# Mesoscale Crystallographic Helicity in Confined Tellurium Quantum Wires

Moniruzzaman Jamal[1,2,3], I K M Reaz Rahman[2,4], Ke Ma[1], Juhyeok Lee[2,3], Karen C. Bustillo[3], Mark Asta[1,2], Matthew P. Sherburne[1], Daryl C. Chrzan[1,2], Ali Javey[2,4,5], and Mary Scott[1,2,3] *

[1]Department of Materials Science and Engineering, University of California, Berkeley, Berkeley, CA 94720, USA

[2]Materials Sciences Division, Lawrence Berkeley National Laboratory, Berkeley, CA 94720, USA

[3]National Center for Electron Microscopy, The Molecular Foundry, Lawrence Berkeley National Laboratory, Berkeley, CA 94720, USA

[4]Electrical Engineering and Computer Sciences, University of California, Berkeley, Berkeley, CA 94720, USA

[5]Kavli Energy NanoScience Institute at the University of California, Berkeley, Berkeley, California 94720, United States

*Corresponding author: mary.scott@berkeley.edu

**Abstract**

**Helical order can facilitate symmetry breaking and emergent physical responses in crystalline materials, yet how intrinsic chirality manifests beyond atomic length scales remains poorly understood. Here, the direct observation and quantitative characterization of long-range crystallographic helicity in template-grown tellurium (Te) quantum wires on amorphous substrates are reported. Four-dimensional scanning transmission electron microscopy (4D-STEM) enables quantitative mapping of crystallographic orientation with nanometer-scale spatial resolution. The resulting orientation maps establish continuous mesoscale lattice twisting, providing direct evidence of long-range crystallographic helicity. Correlated orientation and strain mapping reveal pronounced lateral strain heterogeneity, with compressive strain concentrated within the wire interior. Systematic analysis across multiple wires suggests that higher twist rates are generally associated with weaker lateral compressive strain, narrower wires, and better atomic chain - template axis alignment. Complementary first-principles calculations on finite Te nanorods further suggest that twisting is intrinsically accessible in nucleus-scale Te clusters and strain can bias the preferred torsional state. Together, these results support a growth-incorporated, strain-biased picture in which nanoscale confinement and anisotropic strain facilitate torsional relaxation and stabilize mesoscale helicity in Te nanostructures highlighting strain and confinement as potential routes for engineering chiral lattice states in van der Waals nanostructures.**

## Main

## Introduction

Helical and chiral order in crystalline solids provides an effective route to breaking symmetry and generating new functionality, enabling novel electronic, optical, and thermal responses beyond those accessible in centrosymmetric materials[1–3]. In low-dimensional systems, helicity can couple strongly to elastic strain, confinement, and interfacial effects, giving rise to mesoscale twisting phenomena that are fundamentally distinct from molecular-scale chirality[4]. Such twisted nanostructures have attracted increasing interest for strain-engineered band modulation, chiral charge transport, and mechanically adjustable device designs[5–7]. Tellurium (Te) stands out as a promising platform for exploring emergent helicity across different length scales. Crystallized Te has a trigonal structure composed of one-dimensional helical chains held together by van der Waals interactions, which makes it intrinsically chiral and highly anisotropic[8]. This unique structural anisotropy in nanoscale Te gives rise to diverse functionalities, including anisotropic thermal transport and photocurrent response[9,10]. The lack of inversion symmetry in its chiral lattice also facilitates spin-polarized electronic states and magneto-chiral transport phenomena, as well as circular dichroism[11–14]. These properties have positioned low-dimensional Te as a promising material for next-generation electronics and optoelectronics. Accordingly, understanding the structural intricacies and their impact on the diverse properties of low-dimensional Te is crucial. While the atomic-scale helicity of individual Te chains is well established, how this intrinsic chirality manifests at mesoscopic length scales remains poorly understood.

Twisting in crystalline nanostructures has been attributed to mechanisms including topological defects, growth-environment fluctuations, and intrinsic elastic torques[15–17]. Generally, twisted single crystals arise from growth-actuated, nonuniform stresses that are most efficiently relaxed through torsional deformation in elongated, low-dimensional geometries[18]. Screw dislocation mediated directional twisting, commonly referred to as Eshelby twist, is widely regarded as the dominant mechanism underlying long-range helicity in free-standing nanowires[19]. However, template-grown nanostructures experience a fundamentally different crystallization environment, characterized by in-plane growth, nanoscale confinement, and strong interfacial interactions. Whether these conditions give rise to a distinct mechanism for the emergence of mesoscale helicity remains unknown.

Experimental investigations of chirality in tellurium have so far relied largely on indirect or highly localized probes. Atomic-scale helicity in naturally occurring Te has been imaged using aberration-corrected scanning transmission electron microscopy (STEM)[20]. Scanning tunneling microscopy has served as a powerful tool for probing intrinsic or induced chirality in monolayers and ultrathin nanotubes; however, these surface-sensitive approaches do not capture long-range crystallographic rotation or strain[21,22]. Twisted morphologies in nanowires and related chiral materials have also been reported based on STEM and selected area electron diffraction measurements[23]. However, these techniques generally do not enable quantitative mapping of crystal orientation, which makes it hard to distinguish true lattice rotation from morphological bending, and to quantify the twist rate. Recently, four-dimensional scanning transmission electron microscopy (4D-STEM), which records nanobeam electron diffraction (NBED) patterns across a two-dimensional real-space probe array, has emerged as a promising route to probe long-range helicity in Te nanocrystals[24]. Nevertheless, comprehensive orientation mapping with quantitative twist-rate extraction and correlation to local strain, particularly in substrate-grown and template-grown architectures where interfacial confinement is expected to be critical-remains largely unexplored.

In this work, we report the direct observation and quantitative characterization of mesoscale crystallographic twisting in template-grown Te quantum wires on an amorphous substrate. Using 4D-STEM based crystallographic orientation mapping, we resolve continuous lattice rotation along the nanostructure axis with nanometer-scale spatial resolution. This approach enables unambiguous separation of crystallographic twist from bending or tilting effects and allows direct extraction of the local twist rate, well beyond the scale of individual helical chains. Beyond structural quantification, we investigate the role of elastic strain in relation to the observed twisting. Strain tensor maps extracted from the same 4D-STEM datasets reveal spatially correlated strain gradients along the nanostructures, consistent with a torsional deformation mode. Guided by these experimental correlations, first-principles calculations on finite Te nanorod models show that torsional distortions are energetically accessible in nucleus-scale Te clusters and that axial or radial strain can bias the preferred twist state. This provides a microscopic basis for a growth-incorporated, strain-biased picture in which helicity emerges during early crystallization and is subsequently modified or stabilized by nanoscale confinement and substrate interactions, rather than arising solely from conventional screw-dislocation-mediated twisting. Together, our 4D-STEM measurements and first-principles calculations reveal how intrinsic Te chirality, local strain, and geometric confinement can couple to generate mesoscale crystallographic helicity in substrate-grown nanostructures. This work establishes a quantitative framework for resolving and interpreting crystallographic twisting in low-dimensional materials and suggests new opportunities for engineering chiral lattice states through strain and confinement in van der Waals systems.

**Results and Discussion**

Template-grown Te quantum wires were synthesized directly on amorphous silicon nitride (a-SiN) TEM grids using thermal evaporation as described in the experimental section. A schematic diagram representing the Te deposition and structural evolution process is shown in Figure 1a. A bright-field (BF) TEM image of a few of the Te quantum wires (grown 500 nm apart by design) is shown in Figure 1b. The selected area electron diffraction (SAED) pattern in the inset confirms the crystalline trigonal structure of the synthesized Te wires, with the [0001] direction (c-axis) preferentially aligned along the template direction. This observation highlights the ability of nanoscale confinement to direct crystallographic ordering during growth, consistent with our separate work on orientation-controlled Te nanostructures. Figure 1c presents the BF-TEM image of a single Te wire with visible twisting features along with some amount of surface roughness coming from the Poly(methyl methacrylate) (PMMA) template defined by e-beam lithography. The contrast variation observed here points to the possibility of long-range twist along with other structural features like bending. The high-resolution TEM image in Figure 1d confirms the crystallinity of the synthesized Te wires with lattice fringes presented in the inset. The HRTEM images in Figure S1 resolves the characteristic helical atomic-chain structure of crystalline Te. The Fast-Fourier transforms (FFTs) shown in Figure 1e, from different segments (marked 1-4) of the HRTEM image, provide further evidence of local crystallographic orientation changes along the nanowire. Bragg reflections visible in the cropped FFTs evolve systematically as the analysis window progresses along the wire, while the diffraction features corresponding to the axial direction remain nearly unchanged and close to the [0001] orientation. This behavior is consistent with the SAED patterns, which likewise indicate that the overall crystallographic orientation remains nearly aligned with the [0001] direction along the nanowire axis. The systematic evolution of the non-axial diffraction spots, together with the preserved axial orientation, suggests that the crystal undergoes continuous rotation about the nanowire axis rather than changes in the growth direction or the formation of discrete crystallographic domains.

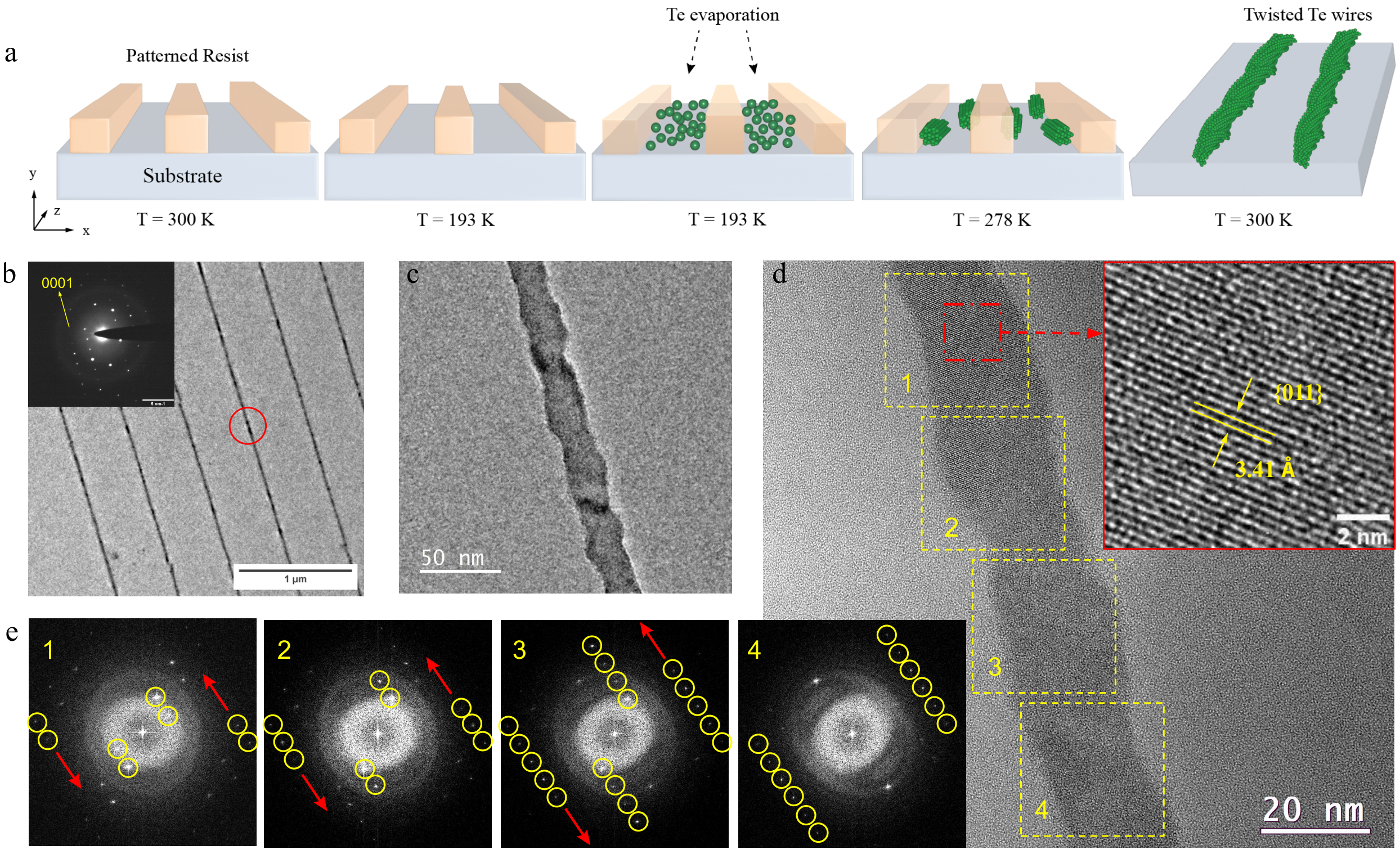

**Figure 1.** Synthesis and structural characterization of template-grown Te quantum wires on an amorphous substrate. a) A schematic diagram showing the synthesis process. For visual clarity, Te deposited on the resist surface during thermal evaporation is omitted from the schematic. b) BF-TEM image of a few template-grown Te wires, inset showing a SAED pattern with indexed [0001] direction. Scale bar in SAED is 5 $nm^{-1}$ c) BF-TEM image of a single Te wire with visible twisting features. d) HRTEM image of a ~20 nm wire. e) Binned FFTs from different areas along the axial direction of the wire in d). Gradual change of the marked reflections (counterclockwise as indicated by the arrow) as we scan from top to bottom position in the Te wire, is an indication of meso-scale helicity.

Tracking the evolution of SAED patterns along the nanowire axis can qualitatively reveal axial lattice rotation over extended length scales, as shown in Figure S2[25]. However, the limited spatial resolution of SAED prevents reliable quantification of local crystallographic rotation and makes it difficult to distinguish continuous lattice twisting from discrete orientation changes. These limitations motivate the use of 4D-STEM, in which a nanobeam electron diffraction pattern is recorded at every probe position during a two-dimensional STEM scan, producing a spatially resolved diffraction dataset. By combining real-space imaging with local reciprocal-space information, 4D-STEM enables quantitative mapping of crystallographic orientation with nanometer-scale spatial resolution, allowing continuous lattice twisting to be directly resolved and quantified along the entire nanowire.

To study the spatial distribution of crystallographic orientations and quantify the twist rate along the wire, we performed automated crystal orientation mapping (ACOM) using 4D-STEM datasets[26]. We constructed a crystal model for trigonal Te and generated a simulated diffraction library spanning the symmetry-reduced orientation space. Using this, we calculated the structure factors and orientation plan for correlation-based matching. Local orientations were obtained by comparing calibrated experimentally detected Bragg vectors with the simulated library and assigning the best-matching orientation at each probe position. This ACOM-based analysis enabled quantitative mapping of crystallographic twist and served as the basis for correlating orientation gradients with strain extracted from the same 4D-STEM datasets.

Figure 2 summarizes the identification and quantitative analysis of long-range crystallographic helicity in a Te quantum wire using 4D-STEM-based ACOM analysis. Figure 2a presents a virtual dark-field (VDF) image constructed from the 4D-STEM dataset by integrating the diffraction intensity within a virtual annular detector placed around the transmitted beam at every scan position. The resulting image provides diffraction contrast with nanometer-scale spatial resolution while preserving the crystallographic information contained in the nanobeam electron diffraction patterns. In Figure 2b, we present the corresponding orientation map obtained from ACOM analysis. The orientation map reveals a continuous evolution of both the out-of-plane orientation (z-axis) and the in-plane orientation (y-axis) perpendicular to the wire axial direction as a function of position along the wire. In contrast, the in-plane orientation along the axial direction (x-axis) remains nearly constant and close to the [0001] direction. The gradual change in orientation along two directions while the axial direction remains unchanged indicates that the observed orientational variation arises from a continuous rotation of the crystal lattice about the axial direction, rather than changes in growth direction or discrete domain formation. Comparing the color evolution in the orientation map and schematic model of Te hexagonal array, we can see that the out-of-plane orientation gradually changes approximately from $01\bar{1}0$ to $2\bar{1}\bar{1}0$ as we move from left to right along the wire length.

To further investigate and quantify this behavior, we utilized the fiber ACOM framework to extract the axial rotation and out-of-plane tilt, as presented in Figure 2c. We have selected the [0001] axis as the fiber direction for this analysis. The in-plane rotation refers to the azimuthal rotation of the lattice around the wire axis and directly reflects crystallographic twist. Meanwhile, the out-of-plane tilt represents the angular relationship between the [0001] fiber axis and the incident electron beam direction, highlighting the local bending and tilt in the sample. In our dataset, the in-plane rotation changes steadily along the length of the nanowire, accumulating ~90° over about 690 nm length. The out-of-plane tilt remains large and relatively uniform, consistent with the near-perpendicular orientation of the wire axis relative to the beam direction while exhibiting minor local variations due to bending or tilt. This clear distinction demonstrates that the dominant structural evolution is intrinsic axial twisting of the crystal lattice.

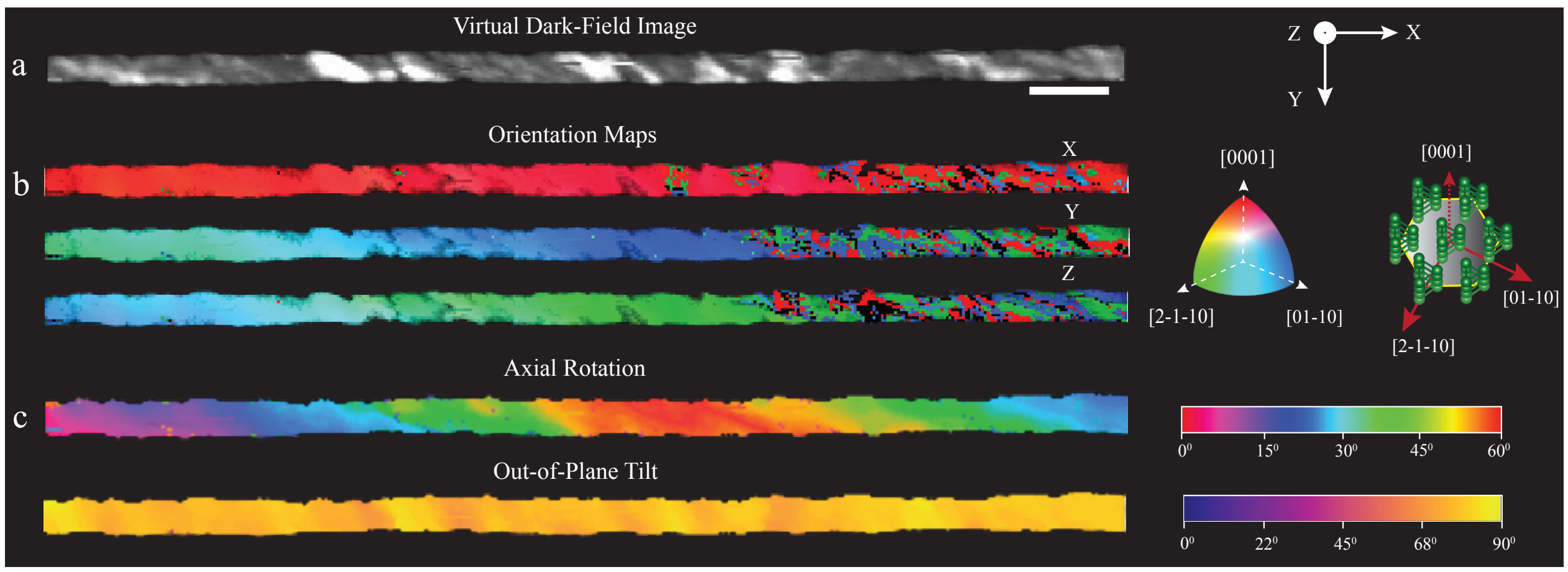

**Figure 2.** 4D-STEM-based orientation analysis for crystallographic twist measurement. a) Virtual dark-field image reconstructed from 4D-STEM dataset (scale bar is 50 nm). b) ACOM-derived orientation maps. The continuous evolution of crystallographic orientation along the nanowire under two orthogonal projections is demonstrating long-range lattice rotation. The persistence of a fixed axial orientation component confirms that the observed variation arises from rotation about the nanowire axis rather than changes in growth direction. Schematic diagram showing the reference orientations in hexagonal array of Te. c) Fiber ACOM maps of axial rotation and out-of-plane tilt with the [0001] direction selected as the fiber axis. The axial rotation exhibits a monotonic progression along the nanowire, directly reflecting crystallographic twist, while the out-of-plane tilt remains large and relatively uniform, indicating that the observed rotation is not caused by crystal bending or tilting.

The schematic diagram presented in Figure 3a helps to visualize the azimuthal rotation angle ($\varphi$) and the tilt angle ($\beta$). Figure 3b provides a complementary visualization of rotational behavior by plotting the ACOM-derived orientation in spherical coordinates. Here, the azimuthal angle ($\varphi$) captures the change of the zone-axis (parallel to beam direction) relative to a reference orientation as we scan along the axial direction of the wire, representing the crystal lattice twist in real space. The polar angle ($\beta$) defines the tilt angle between the zone axis and the axial direction. The systematic color progression around the polar plot demonstrates a continuous change in $\varphi$ with position, further supporting the presence of long-range helicity rather than abrupt orientation switching. Finally, in Figure 3c, we plot the azimuthal angle ($\varphi$) as a function of position along the wire. The value of $\varphi$ varies almost linearly with distance, enabling direct extraction of the crystallographic global twist rate from a linear fit. It is important to note that there are slight variations in the local twist rates along the wire, which can be related to local relaxations or twist stabilization mechanism. This analysis gives us an average global twist rate of 137.89° $\mu m^{-1}$ ($R^2 \approx 0.99$), providing a quantitative, spatially resolved measure of intrinsic long-range lattice twisting derived directly from 4D-STEM dataset.

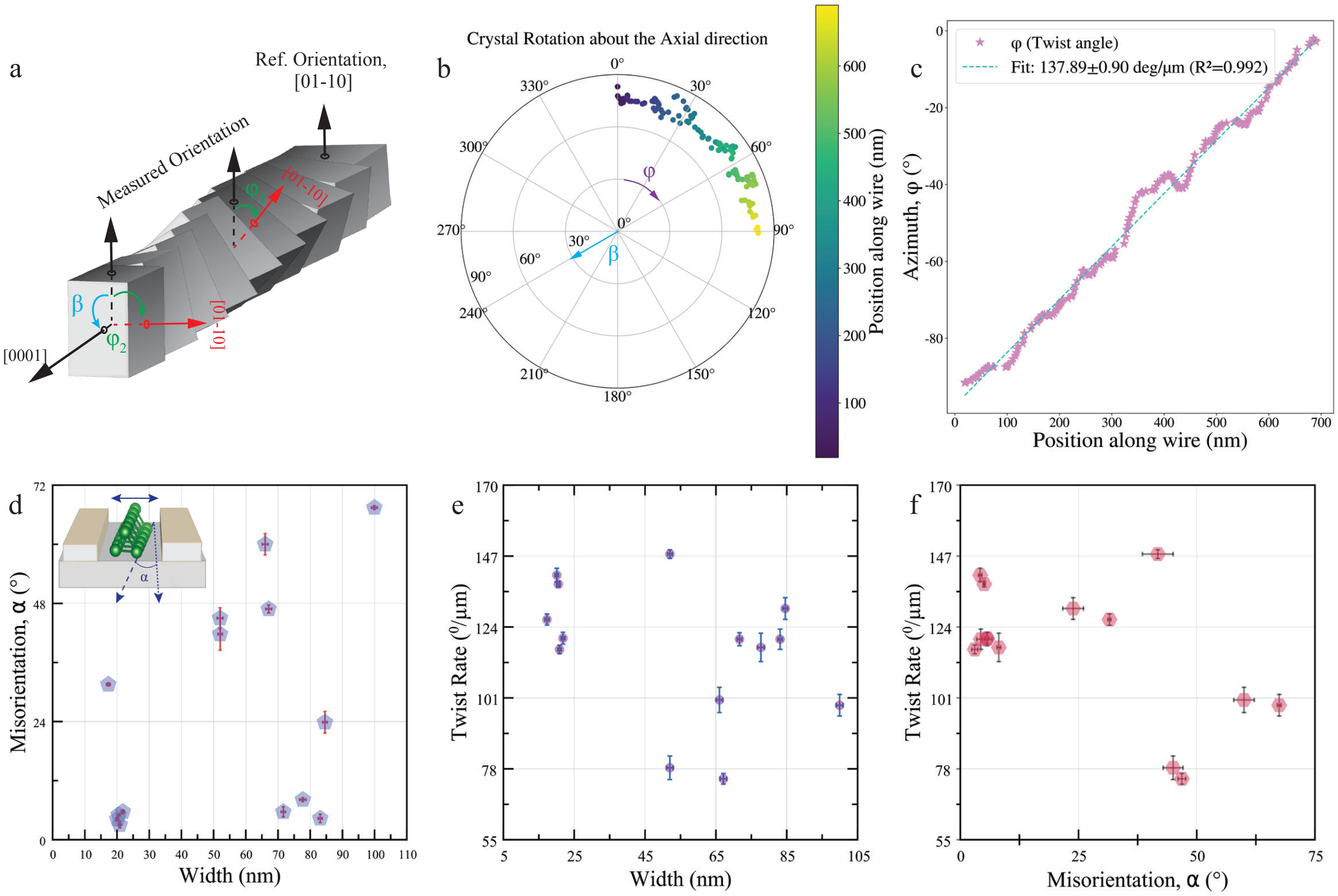

**Figure 3.** Quantification and statistical analysis of twist rate. a) Schematic diagram defining the twist angle ($\varphi$) and tilt angle ($\beta$) in a Te model structure. b) Polar (stereographic) representation of the ACOM-derived zone-axis orientations, color-coded by position along the nanowire, showing a continuous azimuthal rotation about the axial direction consistent with long-range helicity. c) Quantitative extraction of the twist rate obtained by unwrapping the azimuthal angle and fitting its linear evolution with position along the nanowire, yielding a global crystallographic

twist rate. d) Wire (template) width vs misorientation angle between Te atomic chains and template direction, showing the impact of geometric constraint on orientation control. The error bar represents the standard error of measurement (SEM). e) Wire width vs twist rate depicting the influence of geometric parameter on helicity. f) Misorientation vs twist rate showing the impact of crystal anisotropy on the helicity.

To understand what influences the crystallographic twist, we analyzed the dependence of twist rate on geometric confinement and crystallographic anisotropy (Figure 3d-f). We have investigated wire of varying width and crystallographic orientation. We observed that as the width of the wires increases, the orientation of the tellurium [0001] atomic chains becomes more diverse, indicating reduced crystallographic alignment in wider wires. The broader distribution of chain-axis orientations observed in wider wires likely reflects the reduced effectiveness of geometric confinement during crystallization by wider templates. In narrow templates, the PMMA sidewalls impose stronger directional constraints that favor alignment of the Te [0001] chains with the template axis, whereas wider templates permit a larger range of crystallographic orientations. DFT adsorption calculations further indicate appreciable Te-substrate interactions that may reinforce this alignment during nucleation and early growth (Figure S10). Correspondingly, the twist rate decreases with increasing misorientation angle ($\alpha$) between [0001], the chain direction, and the wire axis (template direction), with the largest twist rates observed when the atomic chains are nearly aligned with the template. A similar trend is observed with width: thinner wires exhibit higher twist rates, while wider wires generally twist less. However, in cases where strong chain alignment is preserved, high twist rate is observed even if the width is high.

We quantified the local elastic deformation using ACOM-based strain analysis of the 4D-STEM datasets, enabling direct correlation between crystallographic orientation, strain, and helicity. Figure 4a presents the strain maps generated from the deformation matrices obtained through this analysis. These matrices were constructed by comparing the experimentally measured Bragg disk positions with the corresponding diffraction spots in the simulated patterns of the best-matched orientations determined during ACOM indexing. The rotational component of the deformation matrix is represented by $\theta$, which corresponds to the local crystallographic rotation. The resulting $\theta$ map clearly reveals the continuous lattice twist along the nanowire axial direction. The extracted strain maps reveal a pronounced heterogeneous strain distribution across the nanowire, most notably in the lateral strain component, $\varepsilon_{xx}$. The wire interior is predominantly under lateral compression, whereas the edges show comparatively tensile strain, resembling the core-shell like strain distribution reported in twisted GeS nanowires[3]. However, unlike screw-dislocation-driven twisted nanowires, where strain is typically organized around a central dislocation core[3], our Te nanowires exhibit localized tensile regions within the compressive $\varepsilon_{xx}$ field along the wire length. This spatially non-uniform strain suggests that local relaxation processes may contribute to variations in the local twist rate. More broadly, the coexistence of compressive and tensile regions indicates an elastic state that cannot be fully accommodated by in-plane deformation alone and may instead promote relaxation through out-of-plane modes such as torsion. To further explore the relationship between elastic strain and helicity, we analyzed the correlation between the measured twist rate and the average strain components across multiple wires (Figure 4b-d). Among the examined components, the magnitude of twist rate decreases as the average lateral compressive strain becomes higher, which indicates a strong relationship between residual elastic compression and the degree of mesoscale helicity. In contrast, no clear trend was observed between the twist rate and either the axial or shear strain components. This is perhaps due to insufficient datapoints, although measurable levels of these strains were present in all twisted nanowires.

Researchers have proposed stress-driven twisting mechanisms for a range of low-dimensional crystalline systems. In anisotropic nanostructures, non-uniform surface or edge stresses can generate a net torsional moment, allowing the system to reduce its total elastic energy by adopting a helically twisted configuration rather than accumulating large in-plane strain[4]. A width-dependent twisting behavior driven by elastic anisotropy and confinement has been reported in calcite nanowires, where lattice distortions and shear strain were shown to promote long-range rotation along the wire axis[25]. In the context of Te, intrinsic chirality and highly anisotropic bonding further enhance the coupling between strain and rotation, enabling efficient conversion of strain into torsional deformation even in the absence of a central topological defect[24].

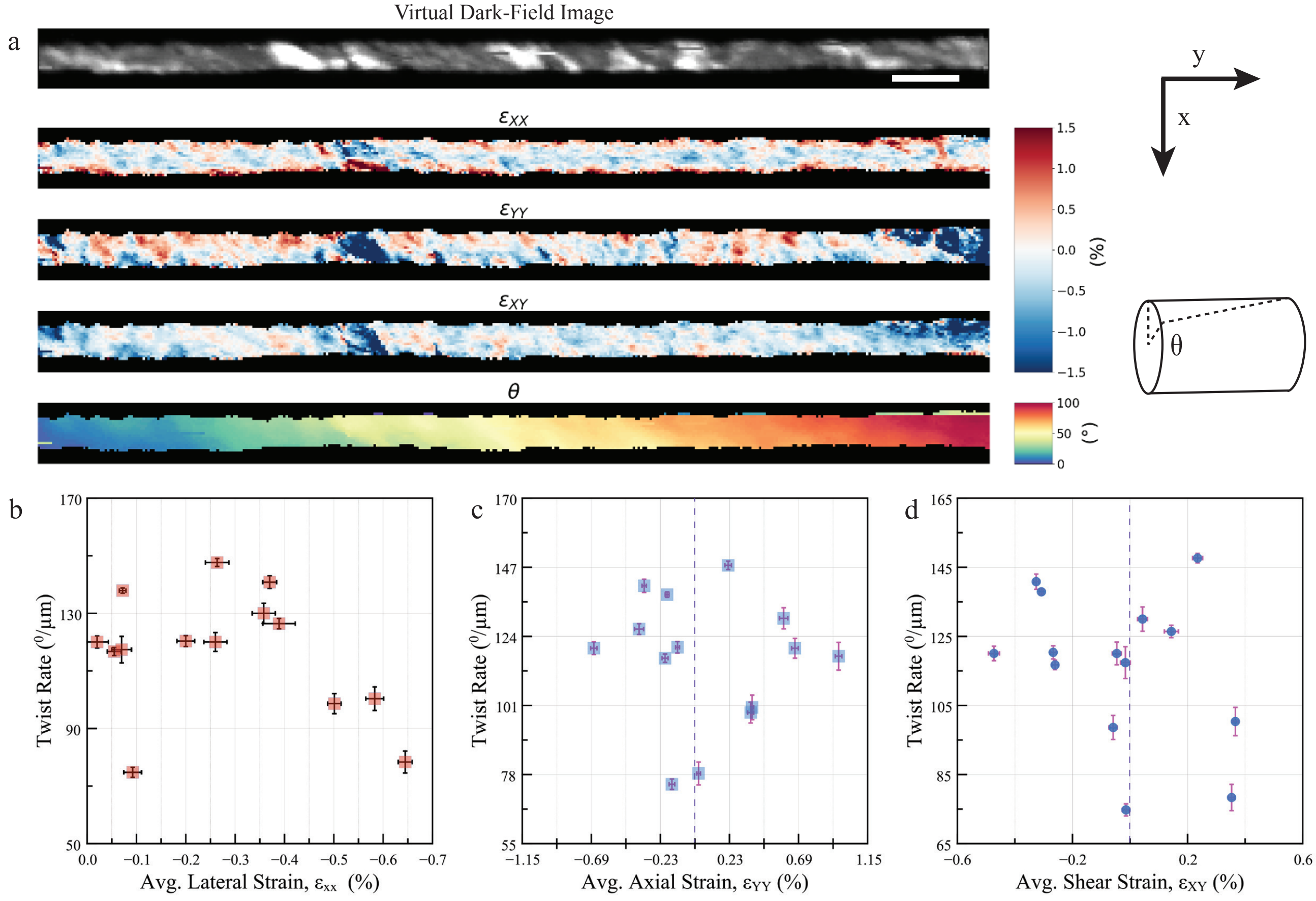

**Figure 4.** Strain mapping and analyzing its impact on twist rate. a) Virtual dark-field images (scalebar is 50 nm), strain maps and spatial distribution of crystal rotation generated using 4D-STEM dataset. b) Average lateral strain vs twist rate. c) Average axial strain vs twist rate. d) Average shear strain vs twist rate.

To investigate the microscopic origin of the experimentally observed crystallographic helicity, we carried out first-principles calculations on finite Te nanorod models designed to approximate the earliest crystalline nuclei formed during confined crystallization. Prior work on low-temperature crystallization of evaporated amorphous Te under similar synthesis conditions estimated a critical nucleus size on the order of approximately 100 atoms[27]. Based on this estimate, we constructed a series of Te nanorod geometries comprising 100-200 atoms as representative models of early-stage nuclei formed from amorphous Te precursor. Full structural relaxation of these hydrogen-terminated finite nanorods consistently produced nonzero spontaneous torsional mode as summarized in Fig. 5a. The relaxed nanorods exhibit average twist rates of approximately $-10\pi$ to $-15\pi$ rad $\mu m^{-1}$ under the sign convention adopted here, in which positive and negative values denote right- and left-handed twist, respectively.

Accordingly, despite Te atomic chains in the starting geometry being exclusively right-handed, all relaxed finite nanorods preferentially develop left-handed torsional states at the rod scale across the segment lengths considered. It was also noted that the local twist is not perfectly uniform along the rod, as reflected by the error bars in Figure 5a. Direct inspection of the relaxed structures reveals a non-uniform, off-center torsional distortion rather than an ideal, rigid-body rotation aligned at the rod's central axis. Nevertheless, results of structural relaxation suggest that a twisted state is energetically favorable in ultrasmall, nucleus-scale Te nanorods even in the absence of external driving force. This observed tendency toward spontaneous twisting implies that helicity may be incorporated at the earliest stages of crystallization.

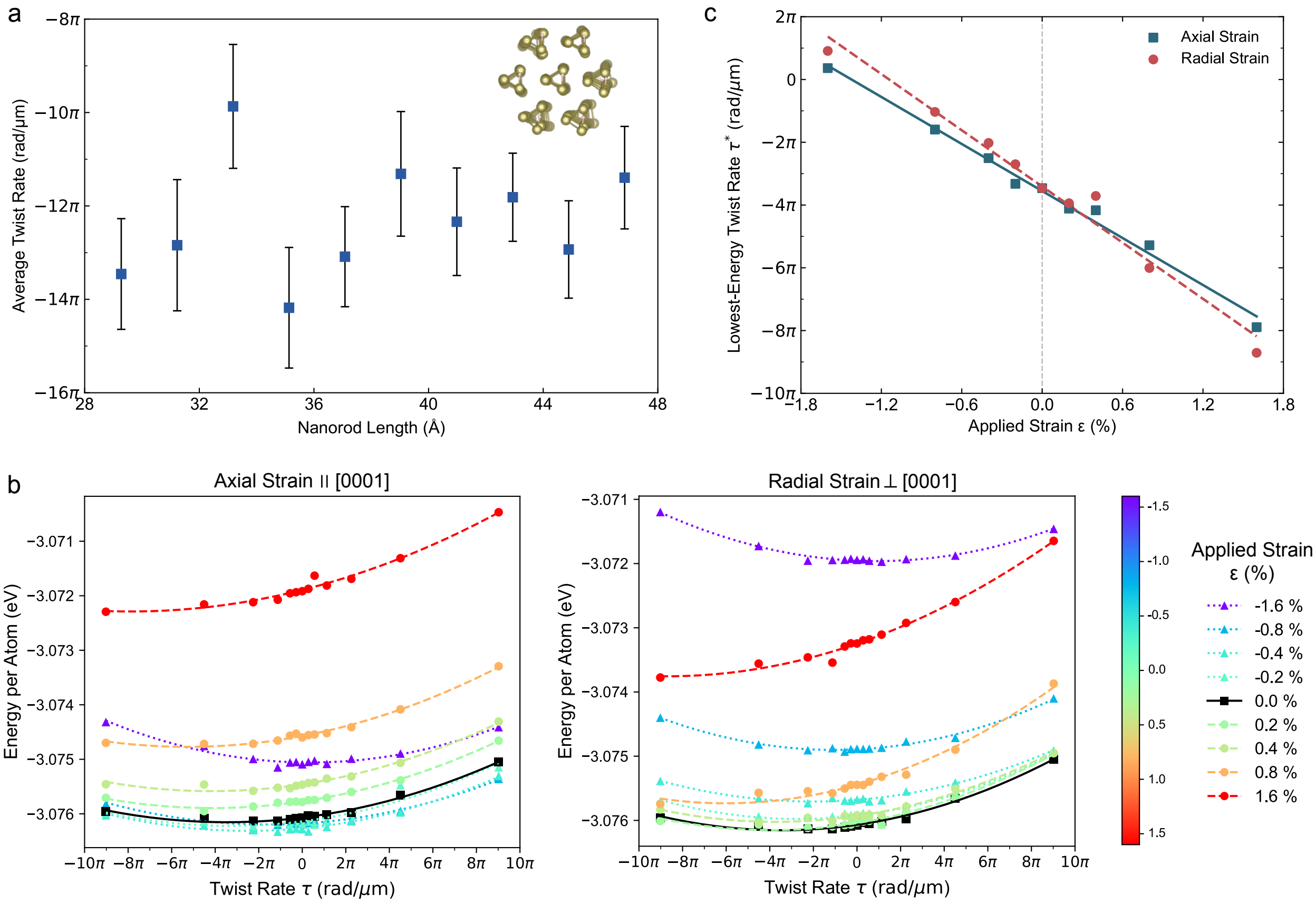

**Figure 5.** Spontaneous twisting and strain-torsion coupling in tellurium nanorods. a) Average spontaneous twist rate as a function of nanorod length obtained from DFT structural relaxations of hydrogen-terminated finite Te nanorod segments. Positive values of twist rate denote right-handed twist, and negative values indicate left-handed torsion. Error bars represent the standard deviation of the local twist rate evaluated along each relaxed nanorod. Inset shows a cross-sectional view of a representative twisted nanorod after relaxation. b) Torsional energy landscapes of strained Te nanorods, showing the energy per atom as a function of imposed twist rate $\tau$ under axial strain applied

along [0001] (left) and equibiaxial transverse (radial) strain applied perpendicular to [0001] (right). Each curve corresponds to a different applied strain $\varepsilon$, with the shared color scale indicating the strain magnitude and sign. Data obtained under tensile strain are shown with circular markers and positive strain values, whereas configurations under compressive strain are shown with triangular markers and negative strain values. c) Energy-minimizing twist rate, $\tau^*$, extracted from quadratic fits to the $E(\tau, \varepsilon)$ curves in panel (b), plotted as a function of applied strain for axial and radial loading.

We then examined whether applied strain can systematically bias the preferred torsional state of chiral Te quantum wires. This question is directly relevant to the experimental environment where 4D-STEM mapping reveals substantial local strain heterogeneity. In addition, DFT adsorption energy calculations (Figure S10) indicate that Te nanowires exhibit notable adsorption strength to amorphous SiN substrate, especially at small nanowire radii. Consequently, any non-uniform substrate interactions could contribute to local structural perturbation and strain buildup in Te nanowires during nucleation and early growth. Strain-torsion coupling at the microscopic scale was modeled using a set of finite nanorod configurations spanning a grid of strain and torsional deformation. Strained states were created by imposing small axial strain along [0001] or equibiaxial transverse strain perpendicular to [0001] (referred to here as radial strain). For each strained state, both right- and left-handed torsions were applied by successively rotating the atomic planes around the nanowire's central axis. The resulting strained-twisted configurations were then evaluated using fixed-ion single-point DFT calculations. This procedure maps the energy of each configuration $E(\tau, \varepsilon)$ as a function of imposed twist rate $\tau$ at fixed applied strain $\varepsilon$, without introducing additional relaxation pathways. To reduce finite-size artifacts associated with the rod ends, the strain-torsion energetics were analyzed using a differential scheme that subtracts total energy of paired long and short nanorod segments with matched terminations, thus isolating the energy contribution of rod interior only.

The resulting torsional energy landscapes are shown in Figure 5b. For both axial and radial loading, energy exhibits a well-defined minimum at finite twist, indicating that the preferred state remains twisted rather than untwisted across the range of strain mode. Moreover, the position of this minimum shifts systematically with applied strain. The corresponding energy-minimizing twist rate, $\tau^*$, extracted from quadratic fits to the local energy profile $E(\tau, \varepsilon)$ at each strain value, is plotted in Figure 5c and varies approximately linearly with both axial and radial strain. These results provide a direct atomistic signature of strain-torsion coupling in ultrasmall Te nanorods by showing that applied strain influences the energetic penalty associated with twisting and determines the twist state that minimizes total system energy. The near-linear dependence of $\tau^*$ on $\varepsilon$ is consistent with a minimal phenomenological description of chiral filaments in which strain and torsion are coupled at lowest order. In such a picture, the energy may be written as

$$E(\tau, \varepsilon) = \frac{1}{2}C\varepsilon^2 + \frac{1}{2}\kappa(\tau - \tau_0)^2 + \lambda\varepsilon(\tau - \tau_0)$$

Where, $C$ is an effective elastic modulus, $\kappa$ is the torsional stiffness, $\tau_0$ is the spontaneous twist at zero strain, and $\lambda$ quantifies strain-torsion coupling. Minimization with respect to $\tau$ at fixed $\varepsilon$ gives

$$\tau^*(\varepsilon) = \tau_0 - (\lambda/\kappa)\varepsilon$$

which naturally yields a linear dependence of the preferred twist with strain. While not intended as a quantitative continuum model, this expression provides a compact framework that captures the trend observed in the DFT energy landscape and highlights the symmetry-allowed coupling between strain and torsion in chiral Te. Notably, analogous twist-stretch coupling has been reported in other chiral quasi-1D systems such as single-walled carbon nanotubes that exhibit axial-strain-induced torsion with equal and opposite rotations in tension versus compression under small strain[28]. More broadly, energy-based theoretical treatments of chiral filaments emphasize that such coupling can emerge generically from chirality and can be approximately linear at small strains[29], consistent with our phenomenological observations captured by the $\lambda\varepsilon\tau$ term.

Taken together, these calculations support a growth-incorporated, strain-biased picture for the origin of helicity in confined Te quantum wires. At the atomistic level, nucleus-scale Te clusters already exhibit a spontaneous twist, while both axial and radial strain systematically bias the energetically favored torsional state. While the calculations are not intended to quantitatively reproduce the full mesoscale mechanics of the experimentally observed nanowires, they identify two components required for a plausible microscopic mechanism: first, twisting is intrinsically accessible in ultrasmall chiral Te nanorods; second, local strain can bias both the magnitude and handedness of the preferred torsion. This picture is fully consistent with the experimental observation of continuous long-range helicity correlated with strain and supports the view that the observed helicity is introduced during nucleation and early growth, then modified or stabilized by the local strain environment during subsequent crystallization.

## Conclusion

In summary, this work presents a direct and quantitative characterization of long-range crystallographic helicity in template-grown tellurium quantum wires. Using 4D-STEM technique combined with ACOM analysis, we resolve continuous lattice rotation along the nanowire axis with nanometer-scale spatial resolution, corresponding to a twist approaching ~90° over ~690 nm and an average twist rate of ~138° $\mu m^{-1}$. By integrating orientation mapping with strain analysis extracted from the same 4D-STEM datasets, we establish a clear correlation between crystallographic twist and local elastic strain. Systematic trends with nanowire width and crystallographic alignment further indicate that the magnitude of helicity is closely linked to geometric confinement and lattice orientation, with thinner and better-aligned wires generally exhibiting higher twist rates. Complementary first-principles calculations further suggest that twisting is intrinsically accessible in ultrasmall chiral Te nanorods and that local strain can systematically bias the energetically preferred torsional state. The calculations reveal spontaneous torsional relaxation in nucleus-scale Te nanorods even in the absence of external driving forces, while both axial and radial strain modify the preferred twist configuration. Together with the experimental observations, these results support a growth-incorporated, strain-biased picture for the emergence of helicity in confined Te nanostructures. This interpretation contrasts with conventional Eshelby-type dislocation-driven twisting models and instead points to a strain-coupled mechanism enabled by confinement and the intrinsic anisotropy of Te. More broadly, the quantitative framework established here provides a powerful approach for resolving crystallographic twisting in low-dimensional materials and offers new insight into how strain and confinement may influence the emergence of chiral lattice structures at the mesoscale. Such strain-structure coupling may open new opportunities for engineering helicity in van der Waals nanostructures with tunable chiral electronic, optical, and magneto-transport properties.

## Experimental Section

### Material Synthesis

Poly(methyl methacrylate) (PMMA) resist was spin-coated onto commercially available amorphous silicon nitride (a-SiN) membranes, each with a thickness of 10 nm, supported by a 200 µm silicon frame. The resist was patterned using electron beam lithography, and the exposed regions were developed in a solution of methyl isobutyl ketone (MIBK) and isopropanol in a 1:3 volume ratio. High-purity tellurium (Te) pellets (99.999%, Sigma-Aldrich) were deposited onto the patterned membranes in a Edwards thermal evaporation coating system equipped with a customized cryogenic stage. Before deposition, the substrates were evacuated to a base pressure of $1\times10^{-6}$ Torr and cooled to -80°C by flowing cold nitrogen gas. Tellurium was then thermally evaporated at a rate of 4 Å/s, with the film thickness monitored in situ using a quartz crystal microbalance. Following deposition, the substrates were gradually warmed to 5°C by introducing a flow of room-temperature nitrogen gas. The as-deposited amorphous tellurium films were subsequently crystallized by maintaining the substrates at 5°C for four hours. After completion of crystallization, lift-off was performed by immersing the silicon-supported membranes in acetone without any additional agitation. The samples were dried in a fume hood after lift-off.

### TEM characterization

#### I. Data Acquisition

For the brightfield and high-resolution TEM (HRTEM) imaging, we used a FEI TitanX 60-300 microscope operating at 300 kV at the National Center for Electron Microscopy facility of the Molecular Foundry, Lawrence Berkeley National Laboratory. For selected-area electron diffraction (SAED) patterns we used a selected area aperture that defined a region with a diameter of ~165 nm. SAED patterns were acquired from different segments of individual quantum wires to confirm the trigonal tellurium crystal structure and to establish the global crystallographic orientation of the wires relative to their long axis.

We acquired the scanning nanodiffraction datasets for 4D-STEM analysis operating the FEI TitanX 60-300 microscope in STEM microprobe mode with an indicated convergence semi-angle of 0.48 mrad and with spot size 10. Such configuration ensured the presence of discrete diffraction disks avoiding any overlap which is essential for accurate disk detection for orientation mapping and strain analysis. Diffraction data were recorded using a Gatan Orius 830 (2k x 2k) CCD detector with an acquisition time of 100 ms per scan position. For the Te quantum wire presented in the main text, we selected a two-dimensional real space scan area of 20x400 pixels with a step size of 1.73 nm, yielding spatially correlated diffraction datasets suitable for quantitative orientation and strain analysis.

## II. Data Analysis

For visualization of helicity and to quantify the twist rate, we reconstructed virtual images from the 4D-STEM nanobeam electron diffraction dataset. Virtual detectors were placed on selected diffraction disk positions, and the diffraction intensity within these reciprocal-space regions was integrated and mapped back to real space. For twist-band contour analysis, two virtual detectors were placed on a specific Friedel pair of Bragg reflections to reconstruct the VDF image. The resulting virtual dark-field images provide the spatial distribution of diffraction intensity associated with the selected reflections, enabling visualization of helical structure and twist-induced contrast modulations along the nanowire for twist band contour analysis (Figure S4).

We performed automated crystal orientation mapping (ACOM) and strain analysis on the 4D-STEM datasets using the open-source py4DSTEM framework and custom codes[30]. First, we used a correlation template matching procedure using a synthetic vacuum probe to detect all the Bragg disk positions and intensities at each probe positions generating a Bragg vector map (BVM) (Figure S6). Then we calibrated the Bragg peaks correcting for diffraction pattern shifts, detector ellipticity, rotational offsets between real and reciprocal space, and reciprocal-space pixel size (Figure S6). Local crystallographic orientations were determined by cross-correlation between the calibrated experimental Bragg vectors with a simulated diffraction library spanning the symmetry-reduced orientation space of trigonal tellurium and finding the best-matching orientation for each probe position (Figure S7)[26]. Using the resulting orientation solutions, we defined the best-matching simulated DPs at each probe positions as reference reciprocal lattices, and local lattice distortions were quantified from the relative shifts of the experimentally measured Bragg disk positions with respect to these references. The in-plane strain tensor components were subsequently extracted by fitting the local reciprocal lattice vectors and computing their deviation from the reference lattice geometry. This combined ACOM-based orientation and strain analysis enables quantitative mapping of lattice rotation and elastic strain from the same 4D-STEM dataset, allowing direct correlation between local strain fields and crystallographic twist.

**Computational Methodology**

Density functional theory calculations were performed using the Vienna Ab initio Simulation Package (VASP 6.4.1) within plane-wave basis[31–33]. Exchange-correlation interactions were described using the generalized-gradient approximation in the Perdew-Burke-Ernzerhof (PBE) functional[34], and core-valence interactions were treated using the projector augmented-wave (PAW) method with PBE-GGA PAW datasets for tellurium (POTCAR labeled "Te")[35]. A plane-wave kinetic energy cutoff of 600 eV and Gaussian smearing of width 0.03 eV for electronic occupancies were used throughout all calculations. Total energy convergence threshold for self-consistent-field calculations was set to $1 \times 10^{-6}$eV. Symmetry was disabled for all nanowire and nanorod calculations to avoid imposing artificial constraints on subtle torsional distortions and small strain-dependent energy differences.

Trigonal bulk tellurium was fully relaxed and used as the structural reference for all subsequent model construction. Bulk relaxation employed a Γ-centered $11 \times 11 \times 11$ K-point mesh, with simultaneous relaxation of the lattice vectors and internal atomic coordinates. A larger periodic parent supercell was then constructed from the relaxed bulk unit cell and relaxed using a Γ-centered $3 \times 3 \times 5$ grid. From this structure, quasi-one-dimensional tellurium nanowire models were generated with the wire axis aligned along the crystallographic ([0001]) direction. Each nanowire cross section consisted of seven coaxial right-handed Te helical chains, with a vacuum spacing of approximately 20Å introduced in the two directions perpendicular to the wire axis. The nanowires were subsequently relaxed using a Γ-centered $1 \times 1 \times 5$ K-point.

To model spontaneous twist in Te nucleus, finite nanorod models were generated by truncating periodic nanowire along the [0001] axis and applying hydrogen termination to end surfaces. The structures were fully relaxed until residual forces were below 0.001 eV $Å^{-1}$. To examine strain-torsion coupling, strained and twisted finite nanorod configurations were generated starting from the nanowire-derived rod geometry. Axial strain was applied along [0001] by scaling the axial coordinates, whereas equibiaxial transverse strain was applied perpendicular to [0001] by simultaneously scaling the two lateral coordinates. To further suppress end-surface contributions in the strain–torsion energetics, we used a differential scheme based on paired long and short finite nanorod segments that differed only by three atomic planes along the axial direction while sharing the same end termination and torsion boundary treatment. For each imposed strain-twist state, the differential energy was defined as $\Delta E(\varepsilon, \tau) = E_L(\varepsilon, \tau) - E_S(\varepsilon, \tau)$, where $E_L$ and $E_S$ are the total energies of the long and short segments, respectively. Because the two segments have nominally identical ends, this subtraction isolates the energetic response of the twisted interior section. Differential energies were reported on a per-atom basis.

**Acknowledgments**

This work was funded by the U.S. Department of Energy, Office of Science, Office of Basic Energy Sciences, Materials Sciences and Engineering Division under Contract No. DE-AC02-05-CH11231 (EMAT program KC1201). Work at the Molecular Foundry was supported by the Office of Science, Office of Basic Energy Sciences, of the U.S. Department of Energy under Contract No. DE-AC02-05-CH11231. We thank Hannah DeVyldere and Stephanie M. Ribet for their support in calibrating scanning nanodiffraction data. We declare that the authors utilized ChatGPT (https://chat.openai.com/chat) for language editing purposes only, and the original manuscript texts were all written by human authors, not by artificial intelligence.

**Author contributions**

MJ and MS conceived the idea for the project and designed the experiments. IKMRR and AJ designed and synthesized the samples. MJ and KCB collected TEM data. MJ processed and analyzed the TEM data with guidance from MS.  KM performed the first-principles calculations and analyzed the results with guidance from DCC, MPS and MA. JL performed the 4D-STEM simulation.  MJ, IKMRR, KM, JL and MS wrote the manuscript. All authors discussed the results and commented on the manuscript.

**Competing interests**

The authors declare no competing interests.

**Data and materials availability**

The data that support the findings of this study are available from the corresponding author upon reasonable request.

## Supplementary Information

# Mesoscale Crystallographic Helicity in Confined Tellurium Quantum Wires

Moniruzzaman Jamal[1,2,3], I K M Reaz Rahman[2,4], Ke Ma[1], Juhyeok Lee[2,3], Karen C. Bustillo[3], Mark Asta[1,2], Matthew P. Sherburne[1], Daryl C. Chrzan[1,2], Ali Javey[2,4,5], and Mary Scott[1,2,3] *

[1]Department of Materials Science and Engineering, University of California, Berkeley, Berkeley, CA 94720, USA

[2]Materials Sciences Division, Lawrence Berkeley National Laboratory, Berkeley, CA 94720, USA

[3]National Center for Electron Microscopy, The Molecular Foundry, Lawrence Berkeley National Laboratory, Berkeley, CA 94720, USA

[4]Electrical Engineering and Computer Sciences, University of California, Berkeley, Berkeley, CA 94720, USA

[5]Kavli Energy NanoScience Institute at the University of California, Berkeley, Berkeley, California 94720, United States

*Corresponding author: mary.scott@berkeley.edu

## Table of Contents

## 1. High Resolution TEM images of the Te quantum wire

Cross-sectional TEM lamellae were prepared using a Thermo Scientific Scios 2 DualBeam FIB/SEM system to investigate the possible presence of a screw-dislocation core within the Te quantum wires. As shown in Figure S1a, no dislocation core is observed in the cross-sectional HRTEM image, providing no evidence for a conventional Eshelby-type screw-dislocation-mediated twisting mechanism. A higher-magnification image of the wire interior (Figure S1b) reveals the atomic arrangement of trigonal tellurium, with the highlighted bonds representing covalent bonding within the helical atomic chain. A top-view HRTEM image acquired along the wire axis is shown in Figure S1c, with FFT presented as inset confirming the single crystallinity of the region. The enlarged view in Figure S1d resolves the characteristic helical atomic-chain structure of crystalline tellurium aligned with the template-defined growth direction.

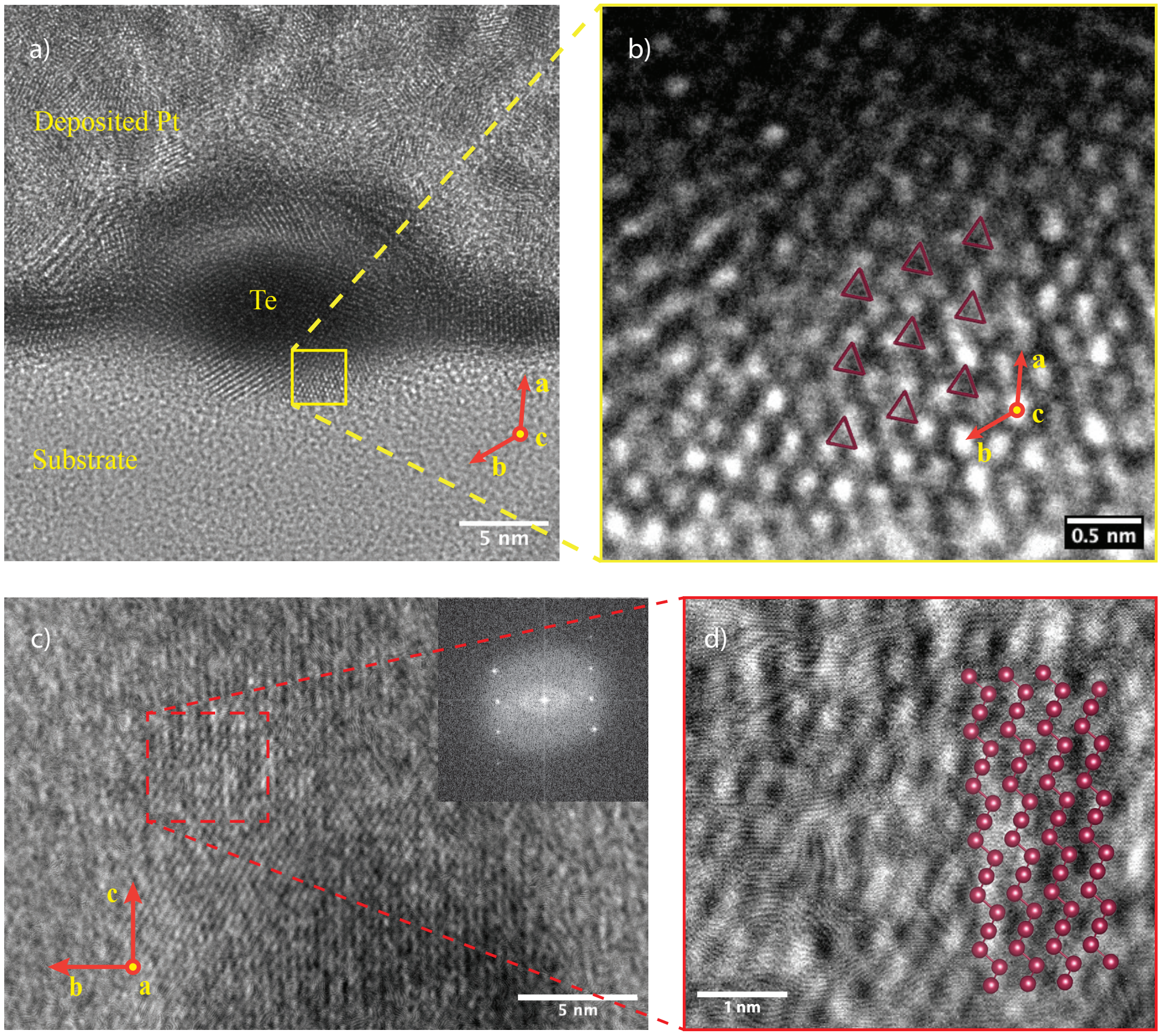

**Figure S1:** High Resolution TEM images of the Te quantum wire. a) Cross-sectional image from a FIB lift-out lamella. b) Magnified view of the marked region showing the atomic arrangements of Te. The overlayed triangular regions represent the covalent bonds in the helical Te atomic chains. c) HRTEM image (top view) of a Te wire with FFT as inset showing the alignment of atomic chains with the template direction. d) Magnified image showing the helical atomic chains. The overlay of Te atomic chains is for visual guide.

## 2. Analysis of SAED patterns for mesoscale twisting

Selected-area electron diffraction (SAED) patterns were acquired using a 10 μm selected-area aperture. Figure S2a shows a bright-field TEM image of Te wires with the aperture locations indicated. The corresponding SAED patterns are presented in Figure S2 (b-d). A gradual evolution of the diffraction pattern is observed as the aperture is translated from the top to the bottom of the wire, while the atomic-chain direction ([0001]) remains unchanged. This behavior provides direct evidence for the presence of long-range crystallographic helicity. However, although SAED can reveal the global twisting behavior of the nanowire, its limited spatial resolution prevents quantitative characterization of local lattice rotation and twist-rate variations along the wire.

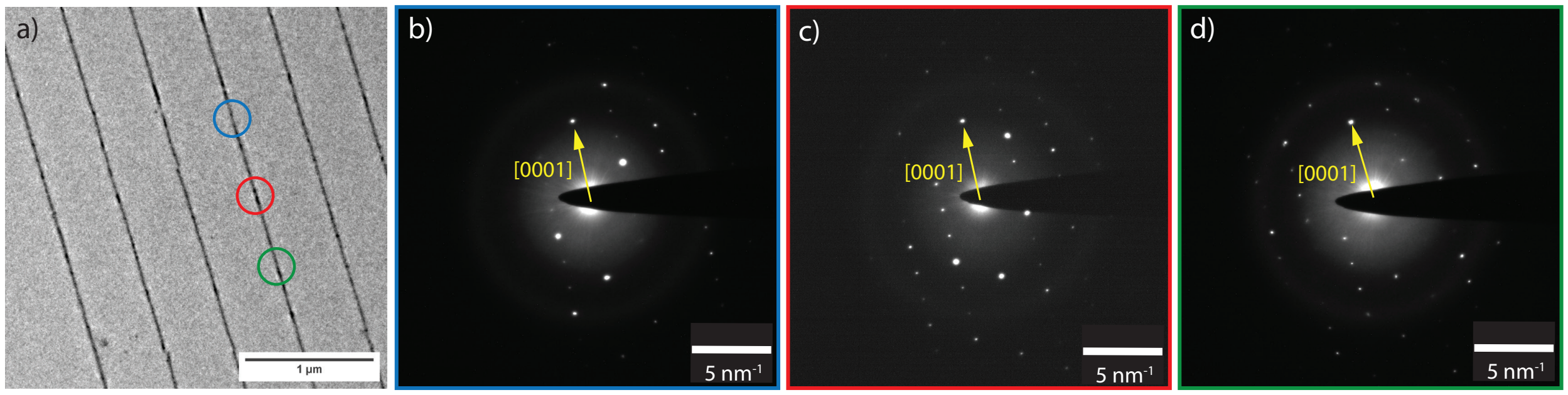

**Figure S2:** Analysis of SAED patterns for mesoscale twisting. a) Bright-field image of template grown Te wires with marked locations of selected area aperture placements. (b-d) SAED pattern corresponding to the marked locations. The gradual change of diffraction patterns while [0001] direction remains unchanged presents the possibility of mesoscale helical twist.

## 3. Class-based reconstruction of virtual dark-field images for meso-scale helicity analysis

We have performed a class-based reconstruction of virtual dark field (VDF) images to study the meso-scale helicity in a Te quantum wire. For this analysis we have used a 4DSTEM dataset collected over 20x180 pixels from a ~25 nm wide wire with a 3.47 nm step size. Placing a virtual annular detector capturing signal from all the diffracted disks, we have constructed the virtual Annular dark-field (ADF) image (Figure S3a). Using py4DSTEM[1], we have analyzed and categorized a total of 3600 NBED patterns in four common classes based on the co-occurrence of Bragg's disks. Placing virtual circular detectors on each of the major disks and summing their signals for each class of NBED patterns, we have constructed the virtual DF images presented in Figure S3(b-e). The reconstructed color-categorized DF image, each color representing a separate class of NBED patterns, shows gradual change of color indicating the crystal rotation about the axial direction. The discontinuity between classes is not real segment, rather coming from weak detection and classification NBED pattern into separate classes. Direct indexing of each class is a bit difficult as some of them are coming from off-zone-axis region. But we can see the clear changes in the reciprocal lattice relrods satisfying Bragg's condition in between classes, consistent with a rotation of the unit cell. We can quantify the twist rate using the last two classes of NBED patterns (yellow & blue). We've indexed the primary vectors in NBED pattern Figure S3i as $g_1 = \bar{2}10$ and $g_2 = 003$, which indicate the zone axis is near $<4\bar{5}10>$. As we move rightward along the axis of Te wire, class NBED pattern changes to Figure S3j, where $g_1 = \bar{3}10$ and $g_2 = 003$ indicating the zone axis to be near $<5\bar{7}20>$. This gives a crystal rotation of ~ $5.2^0$ over the class separation of 50nm in real space image, giving us a twist rate of $0.104^0$/nm for this specific wire.

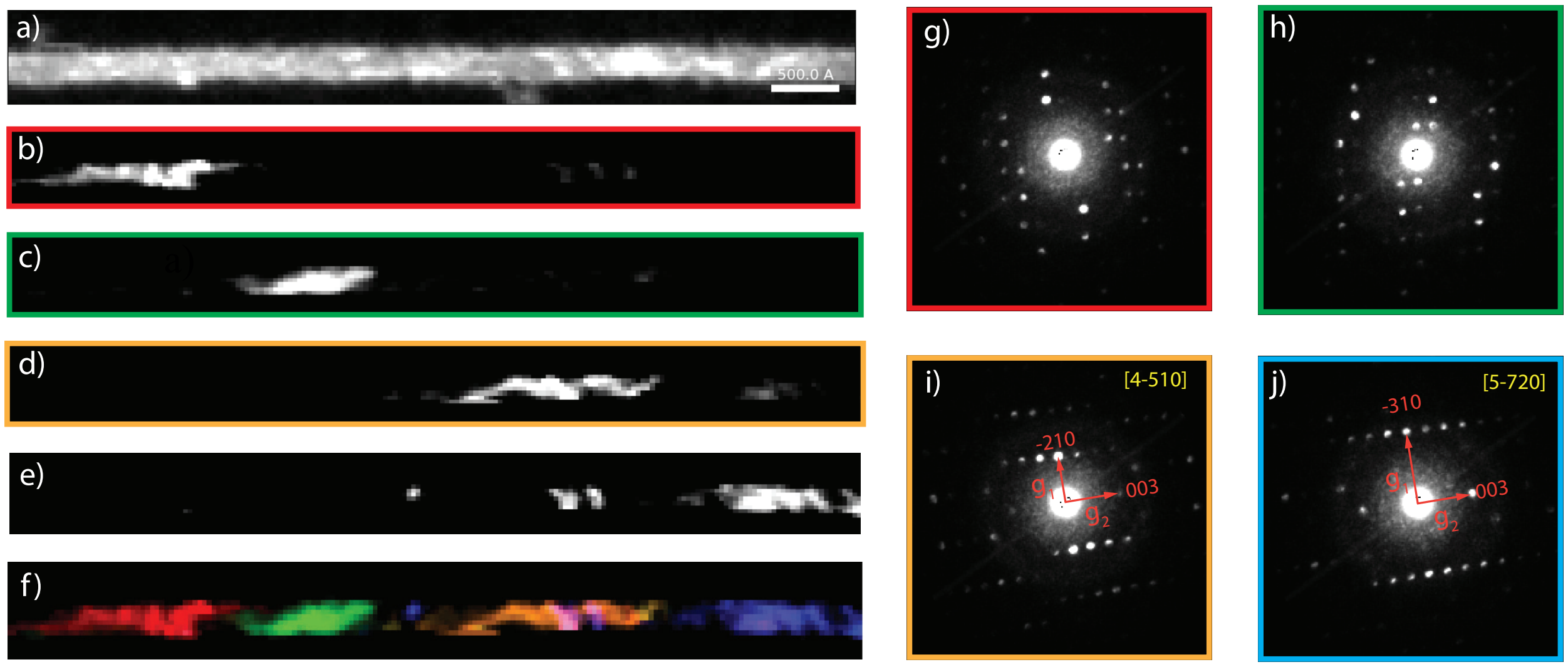

**Figure S3.** Class-based reconstruction of virtual dark-field images for meso-scale helicity analysis. a) Virtual ADF image constructed by placing a virtual annular detector. (b-e) Virtual DF images from contributions of only the major disks of the common classes of NBED patterns. f) Reconstructed color-categorized real space image where each color represents a class of common diffraction patterns. (g-j) Common classes of NBED patterns.

## 4. Twist contour analysis to quantify the twist rate

To measure the twist rate, we have used twist contour analysis using a different 4DSTEM dataset. When the crystal lattice twists along the nanowire growth axis, the reciprocal lattice undergoes a corresponding rotation. Consequently, reciprocal lattice vectors oriented perpendicular to the nanowire axis sequentially satisfy and depart from the Laue condition at different orientations, leading to the formation of multiple twist-induced diffraction contour bands. As depicted in Figure S4, by placing virtual detectors separately on the correct Friedel's pair ($g$ vectors), we can construct virtual DF images and measure the shift of contour bands (L) in real space. Knowing the real space distance of two twist contours, we can calculate the real space twist ($\alpha$) of the wire using the following relation [2].

$$\alpha = (\frac{\lambda}{2L}) \left|\frac{g_2 - g_1}{\sin \beta_g}\right|$$

Where, $\lambda$ is the wavelength of e-beam (1.9 pm for 300kV), $g_1$ and $g_2$ are the $g$ vectors and $\beta_g$ is the angle between $g$ vectors and the axial direction. As shown in Figure S4, we get a twist band shift of ~ 8nm selecting (420) & $\bar{4}\bar{2}0$ as $g$ vectors with $g_2 - g_1 = 1.953\ A^{-1}$ and $\beta_g = 75^0$. This gives us a twist rate of ~$0.14^0$/nm for this sample which seems very reasonable.

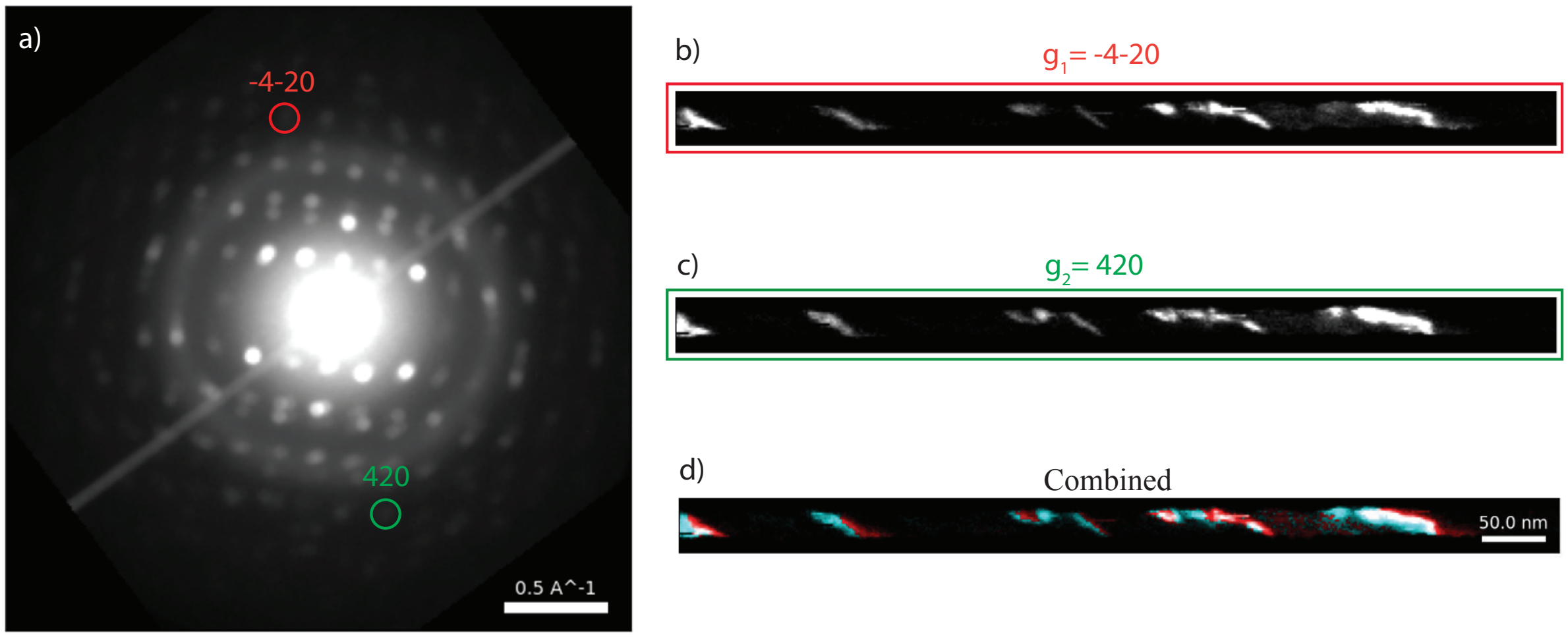

**Figure S4.** Twist contour analysis to quantify the twist rate. a) Mean diffraction pattern from 4DSTEM dataset with marked position of virtual detectors for virtual DF image reconstructions. b-c) Virtual DF images reconstructed using the g vectors. d) virtual DF images overlayed on top of each other showing the shift of the twist contours.

## 5. Comparing experimental data with the simulated 4DSTEM data of a twisted Te quantum wire

The virtual annular dark-field (ADF) STEM image of a Te quantum wire (width ~22 ± 2 nm) reconstructed from a 20x400 pixel array of 4DSTEM dataset is presented in Figure S5a. Utilizing the twist contour analysis, which was originally described by Drum, we have measured the twist rate from a 4DSTEM dataset[3]. Using the measured twist rate of $\sim 0.14^0$/nm, we have modeled a Te nanostructure closely matching the experimental dimensions (Figure S5b) and performed 4DSTEM simulation using abTEM[4]. The diffraction patterns (Figure S5c) corresponding to the marked spots in the virtual DF image gradually changes as we shift from left to right. This phenomenon is clearly captured in the Supplementary video. Similar shift is also observed in the simulated DPs (Figure S5d) obtained from the twisted modeled structure validating the presence of long-range twist in the quantum wire. Here, we note that the key point is that the overall trend in the spot shifts observed in the experimental dataset is consistent with that in the simulated diffraction patterns, even though the positions do not match exactly point by point. This discrepancy likely arises from slight local tilts and distortions in the real sample, which can significantly affect the diffraction patterns in the Te system. Similar axial lattice rotation was previously report for Te NWs with diameter <10 nm [5]. However, NWs with larger diameter and nanoribbons did not show any significant changes behaving just like 2D Te thin films. In the present study of template-grown Te quantum wires, continuous lattice rotation was evident in wires of different width ranging from ~17 nm to ~110 nm.

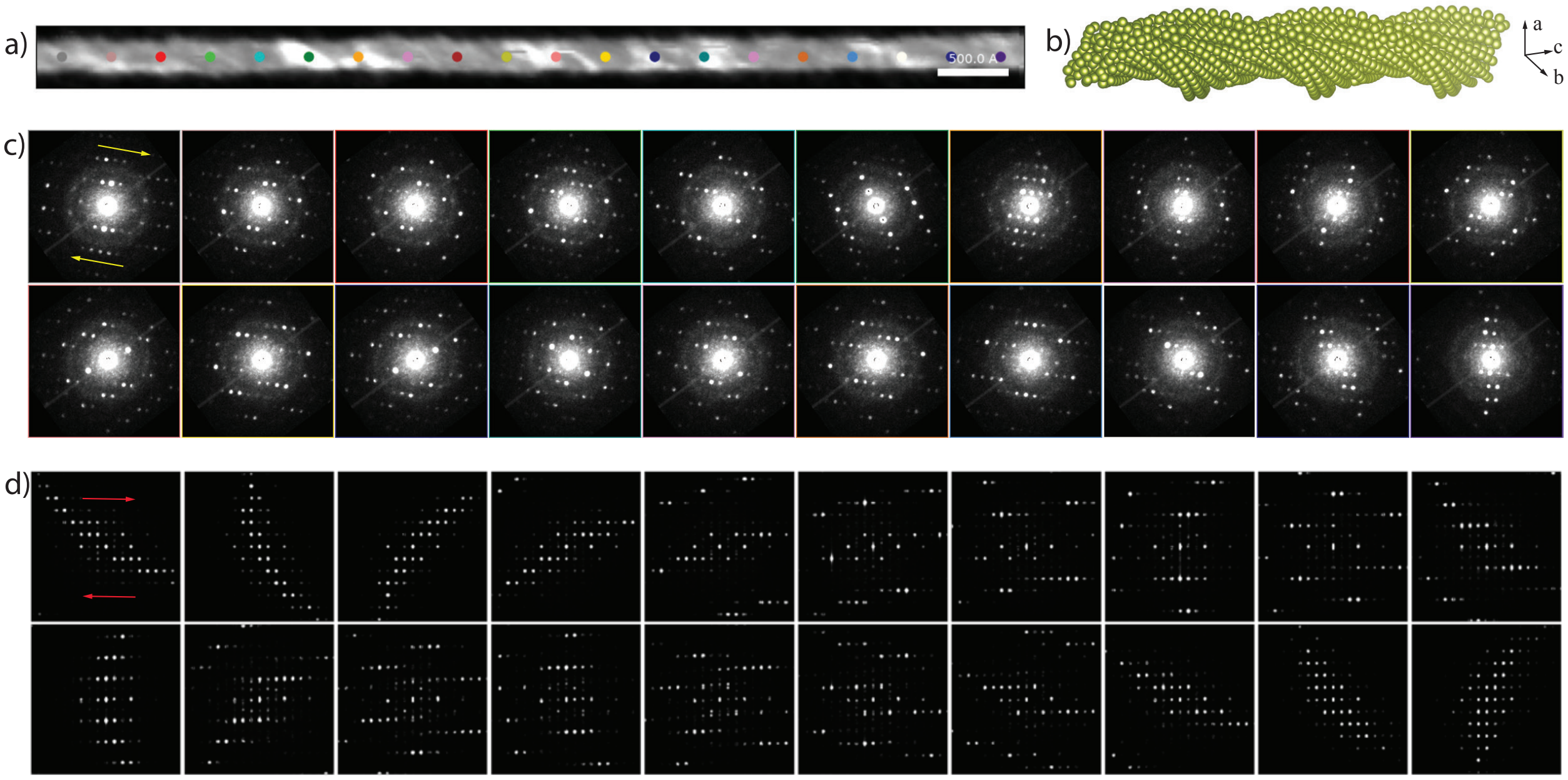

**Figure S5.** Validating mesoscale crystal rotation by comparing with simulated 4DSTEM data. a) Reconstructed virtual DF image of Te quantum wire. b) Schematic of the model twisted Te nanostructure used for abTEM based 4DSTEM simulation. c) Diffraction patterns corresponding to the marked locations of DF image. Direction of the gradual shift of the diffraction spots is indicated by yellow arrows. d) Simulated diffraction patterns obtained from twisted Te nanowire using abTEM simulation. Gradual rotation of diffraction spots (red arrow direction) also observed in the simulated NBED patterns consolidates the presence of continuous lattice rotation in our quantum wire.

## 6. Calibration of 4DSTEM dataset for ACOM analysis

We performed automated crystal orientation mapping (ACOM) on the 4DSTEM datasets using the py4DSTEM analysis framework (Figure S6)[6]. We first identified the positions of diffracted Bragg disks at each probe position using a correlation-based template matching approach, with a synthetic vacuum probe used as the reference template. This procedure yielded the positions and intensities of all detected Bragg reflections across the dataset. The spatial distribution of these Bragg vectors is shown by constructing a 2D histogram, referred to as the Bragg vector map (BVM), in Figure S6c.

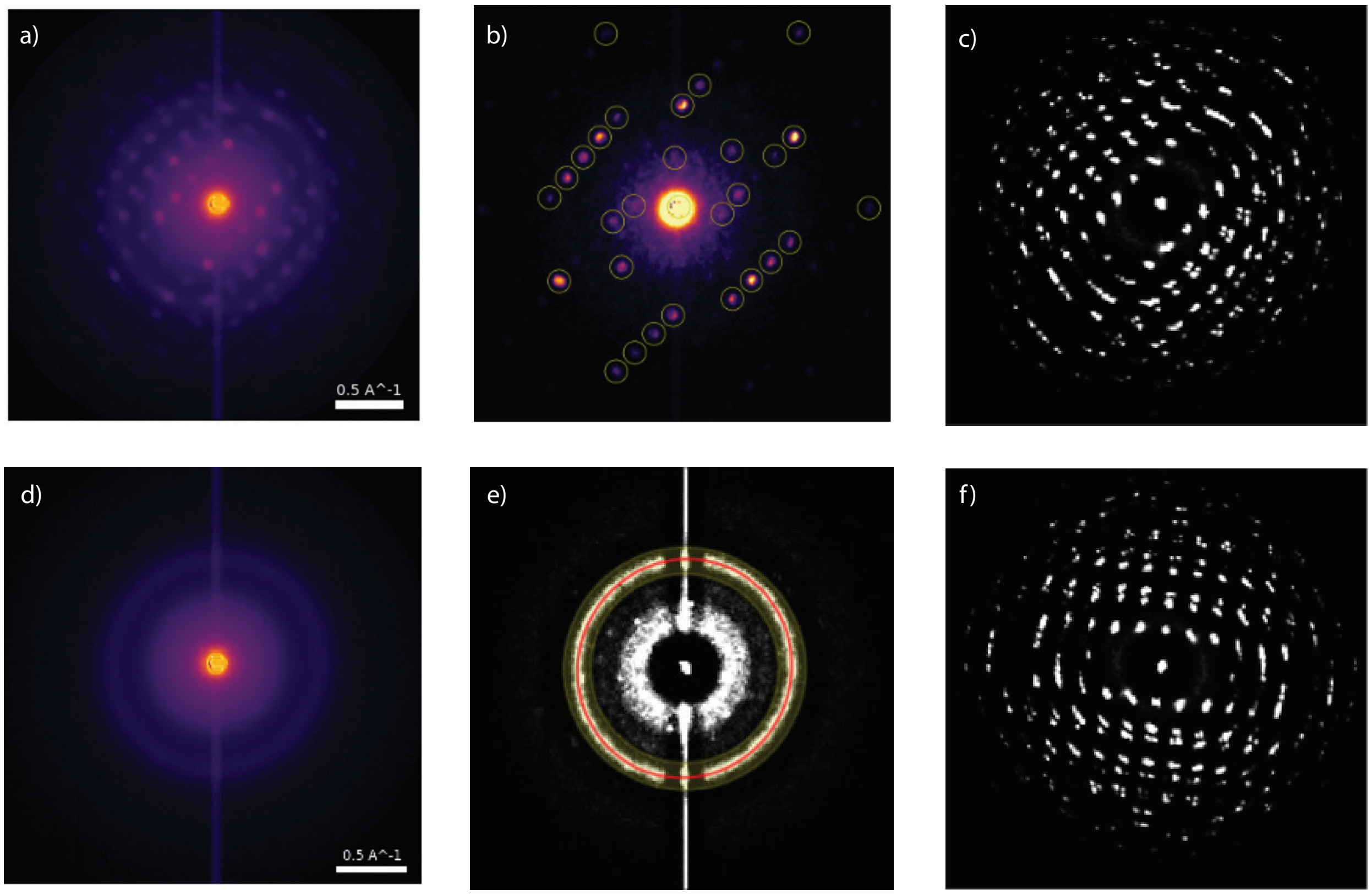

**Figure S6.** Calibration of 4DSTEM dataset. a) Mean diffraction pattern of all the NBED patterns. b) Disk detection by template matching. c) BVM presenting all the detected disk locations. d) Mean diffraction pattern from the amorphous region. e) Ellipticity measurement from the halo ring. f) Calibrated BVM for orientation and strain mapping.

Because experimental 4DSTEM datasets typically exhibit distortions such as diffraction pattern shifts and ellipticity, we subsequently centered the diffraction patterns and corrected for elliptical distortions. We performed the ellipticity calibration by fitting an ellipse to the diffuse halo ring in the mean diffraction pattern obtained from amorphous regions of the dataset, as fitting an ellipse to the single-crystalline tellurium diffraction pattern can introduce systematic errors. As final preprocessing steps, we calibrated the absolute reciprocal-space pixel size and the relative rotation between the diffraction and real-space images. The resulting calibrated Bragg vector map is shown in Figure S6f.

## 7. Cross-correlation method of comparing experimental Bragg peaks with simulated diffraction library

At each probe position, we used cross-correlation approach of comparing the calibrated experimentally detected Bragg peak positions and intensities with a simulated diffraction library generated from the tellurium crystal structure. We first identified a reduced subset of candidate orientations by matching the magnitudes of the experimental Bragg vectors $|g_{exp}|$ to those in the library. We then refined these candidates by accounting for in-plane rotations of the simulated diffraction patterns and computed a correlation score (C) for each orientation, which quantifies the agreement between experimental and simulated Bragg peak positions and intensities (Figure S7). The orientation that maximized this correlation score was selected as the best match, while a correlation threshold was applied to suppress false-positive solutions arising from noise or weak diffraction signal. Although symmetry-related degeneracies may lead to multiple equivalent absolute orientation solutions, the relative orientation changes extracted from the spatial evolution of the best-matching solutions remain robust.

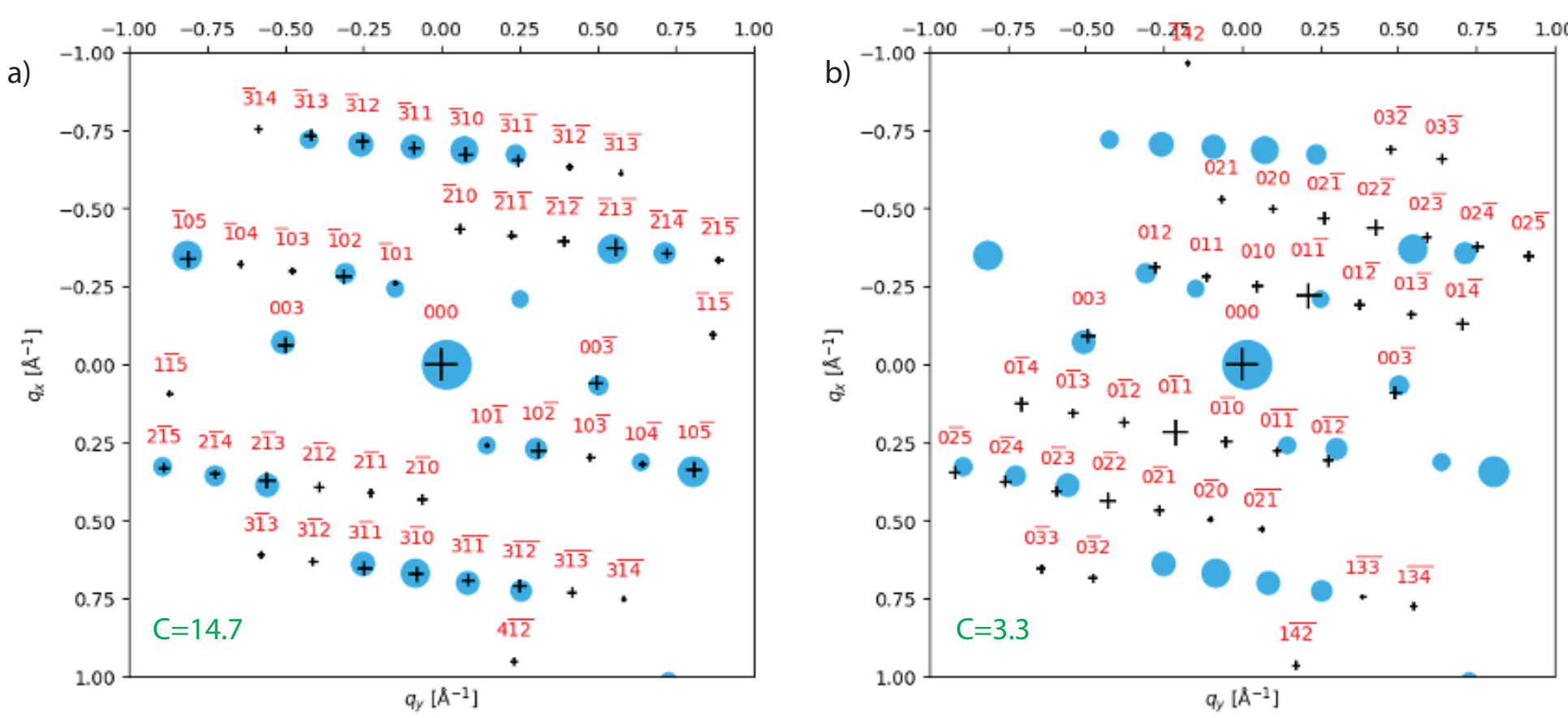

**Figure S7.** Cross-correlation method of comparing experimental Bragg peaks with simulated diffraction library to find the best matching orientation at each probe position. (a-b) Two potential matches for this specific probe position where higher correlation score (C) represents a better match. Slight distortion between the experimental and simulated peaks represents the local strain in the lattice.

To extract lattice strain, we used the ACOM-derived orientation at each probe position to define a local reference lattice. As the crystal planes satisfying Laue conditions for diffraction changes due to continuous crystal rotation, selecting a common pair of g-vectors as reference for strain analysis is not possible for our case. As shown in Figure S7, we compared the local reciprocal lattice vectors $g_{exp}$ (blue dots) obtained from the experimental dataset to the reciprocal lattice vectors $g_{ref}$ (black crosses) predicted by the corresponding best-matching simulated diffraction pattern. The elastic lattice strain was then quantified from the relative shift $\Delta g = g_{exp} - g_{ref}$, which represents local distortions of the reciprocal lattice. By fitting these reciprocal-space distortions across multiple Bragg reflections, we extracted the in-plane strain tensor components. This approach enables simultaneous, self-consistent mapping of crystallographic orientation and strain from the same 4DSTEM dataset, allowing direct correlation between lattice rotation, elastic deformation, and helicity.

## 8. Local relaxation of lateral strain

The lateral strain map obtained from the 4DSTEM analysis is shown in Figure S8a. A heterogeneous strain distribution is observed across the nanowire cross-section, with the outer regions exhibiting predominantly tensile strain while the central region remains largely under compression. The average strain profile extracted from the marked region is plotted in Figure S8b. This alternating strain distribution is indicative of local strain-relaxation processes that facilitate elastic energy minimization within the nanowire. The resulting spatial variation in strain may contribute to the non-uniform local twist rates observed along the wire.

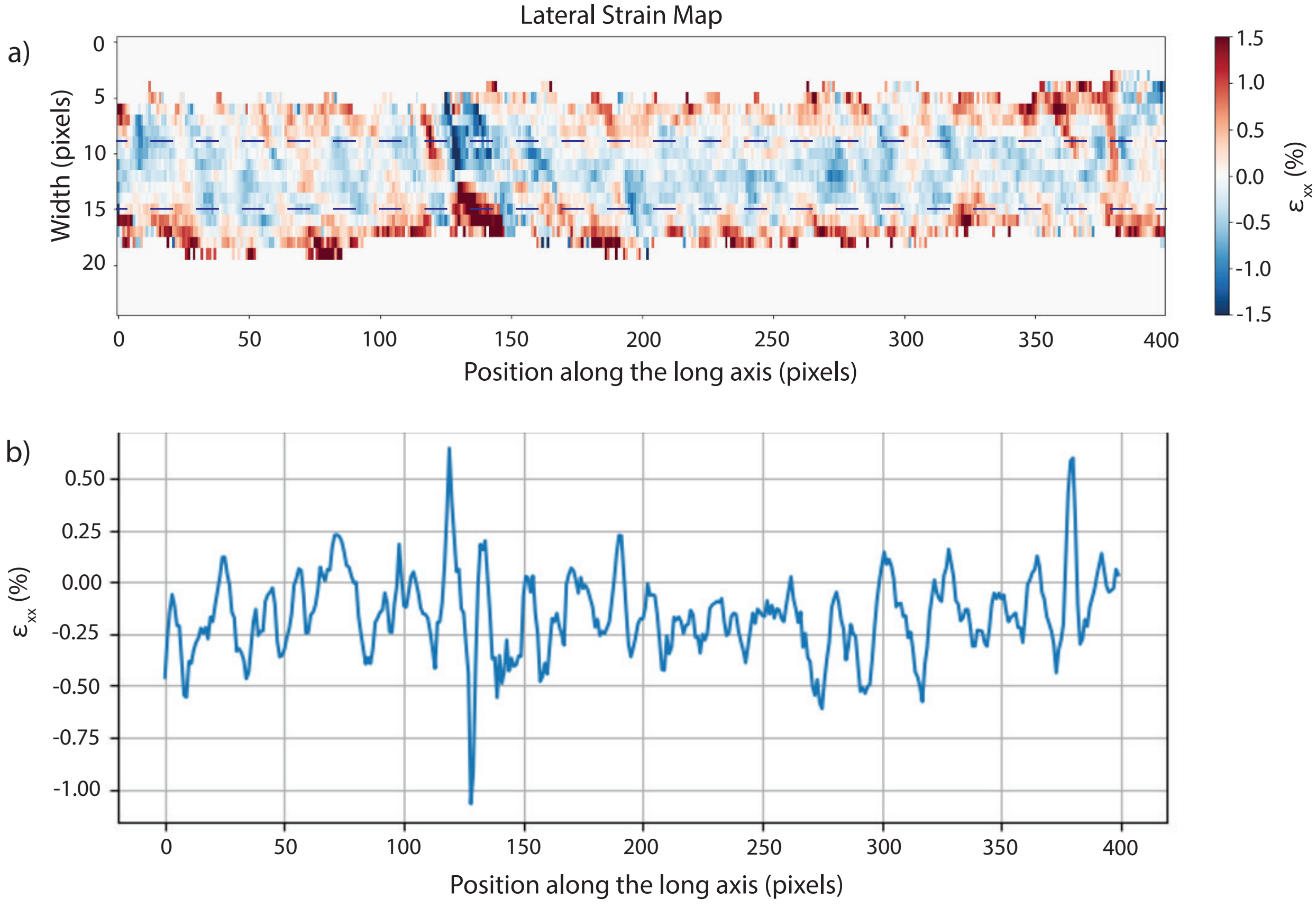

**Figure S8.** Local relaxation of lateral strain. a) Strain map showing the distribution of strain along the lateral direction (x-axis). The size of each pixel is 1.73 nm here. b) Local variations of the strain averaged over the marked region shown in a). Heterogeneity of strain distribution represents the local relaxation of the generated strain during the growth process.

## 9. Different Te quantum wire presenting similar helical behavior

To evaluate the generality of the proposed 4DSTEM analysis framework, we performed the same orientation and strain mapping procedure on an additional Te quantum wire distinct from the representative example discussed in the main text. Here, the step size for 4DSTEM data collection scan was 3.47 nm and the virtual dark-field image reconstructed from the dataset is presented in Figure S9a. The resulting orientation maps (Figure S9b) reveal a continuous evolution of crystallographic orientation along the wire axis, consistent with the presence of long-range lattice rotation. Quantitative analysis of the extracted orientation data reveals a cumulative crystallographic rotation of approximately 165° over ~1400 nm, corresponding to an average twist rate of ~120° $\mu m^{-1}$. The corresponding strain maps (Figure S9c) exhibit heterogeneous strain distributions across the wire cross-section, including regions of lateral compression and tension comparable to those observed in the main dataset. Although the magnitude of the twist and local strain distribution differ from wire to wire, the persistence of continuous lattice rotation across multiple independently analyzed structures demonstrates that the observed helicity is not unique to a single nanowire. These results further validate the robustness of the 4DSTEM-based orientation and strain mapping framework for quantifying long-range crystallographic helicity in template-grown tellurium quantum wires.

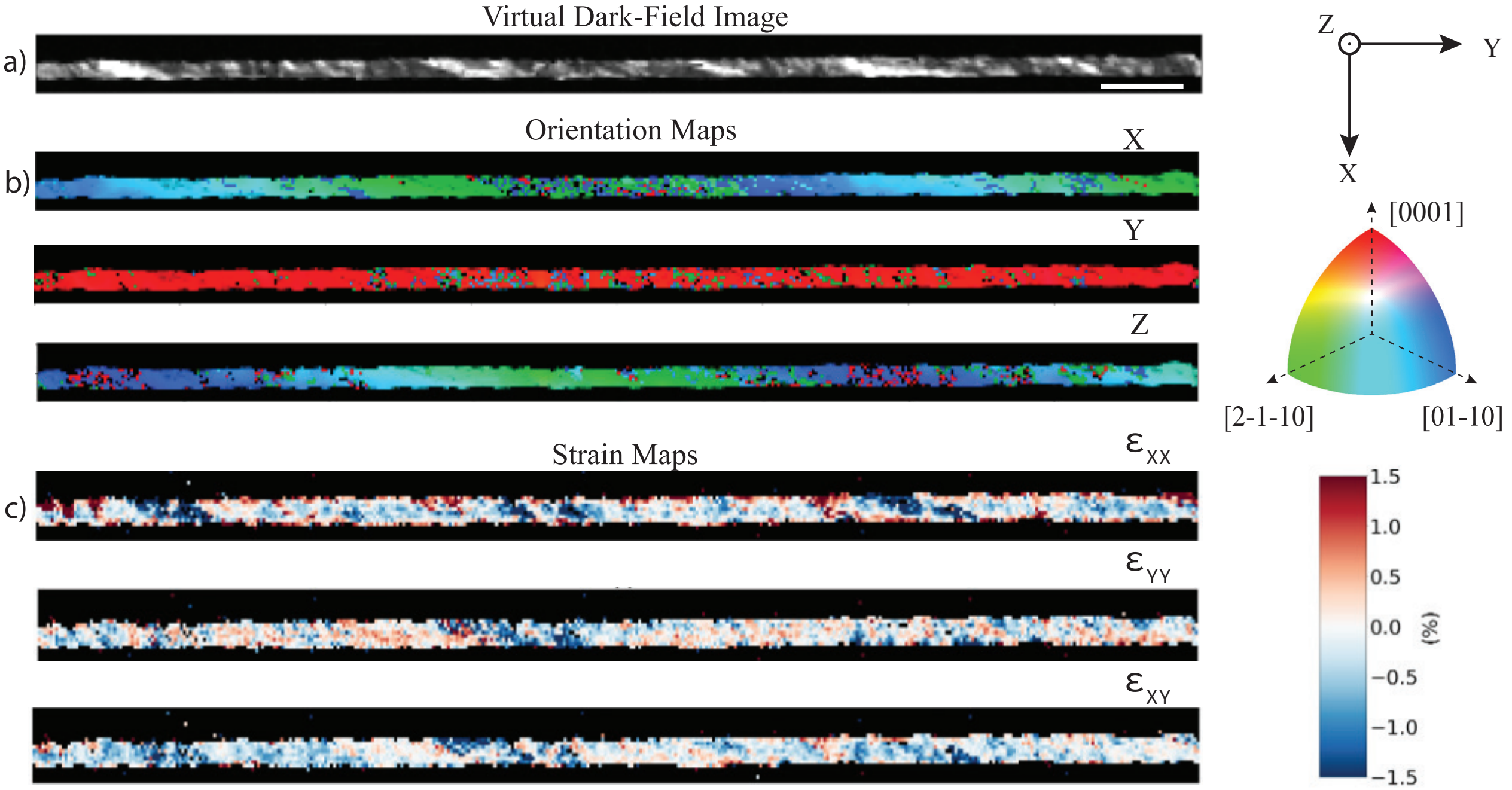

**Figure S9.** Different Te quantum wire presenting similar helical behavior. a) The virtual dark-field image generated from 4DSTEM dataset. The scale bar is 100nm here. b) Orientation map depicting the twisting stricture of the wire. c) Strain maps showing the nature of strain distribution in the confined Te quantum wire.

## 10. Adsorption energy of representative Te configurations on PMMA and amorphous silicon nitride

To investigate the origin of the broad distribution of nanowire misalignment angles observed in the experiment, we evaluated the relative adsorption strength of representative Te building blocks to the two bounding materials present during recrystallization-PMMA sidewalls and the amorphous SiN substrate. Since the initial Te deposit is amorphous and confined within polymer-defined canals during the earliest stages of crystallization, the interfacial energetics at the Te-PMMA and Te-SiN boundaries are expected to promote heterogeneous nucleation and constrain subsequent growth trajectories. We therefore computed the adsorption energies of three

Te configurations spanning increasing structural complexity (dimer, single helical chain, and minimal multi-chain nanowire) on isotactic and syndiotactic PMMA, as well as on H- and OH-terminated amorphous SiN surfaces. The adsorption strengths are summarized in Figure S10.

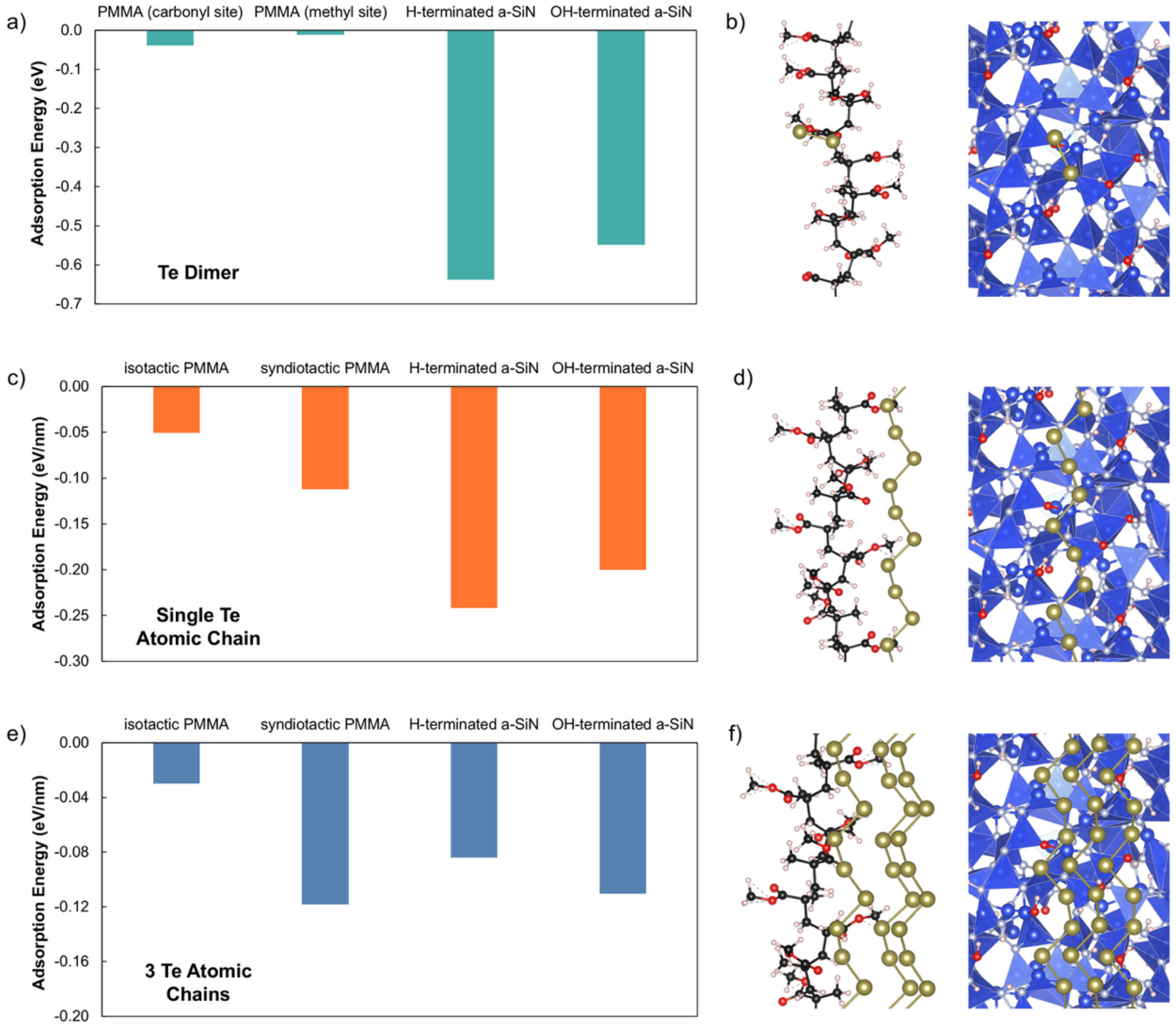

**Figure S10. Adsorption energy of representative Te configurations on PMMA and amorphous silicon nitride.** a) DFT calculated adsorption energies and b) schematic illustrations of the relaxed adsorbed configurations of Te dimer on PMMA polymer chains (left) and OH-terminated amorphous SiN surface (right). c) Length-normalized adsorption energies and d) corresponding relaxed adsorption geometry of single helical Te chain on PMMA (left) and OH-terminated amorphous SiN (right). e) Length-normalized adsorption energies and f) adsorption geometry of 3-chain Te strand on PMMA (left) and OH-terminated amorphous SiN (right). In all bar charts, the vertical axis reports adsorption energy (per stated unit) and the horizontal axis label the adsorption sites.

Across all Te motifs, Te binds more strongly to amorphous SiN than to PMMA, indicating that heterogeneous nucleation is energetically favored at the substrate rather than at the polymer walls. This preference is most pronounced for the smallest Te dimer, suggesting that early nuclei are more likely to be stabilized at the SiN interface than by PMMA template. While adsorption on PMMA remains exothermic in all the configurations examined, the weaker binding implies that PMMA primarily acts as a geometric confinement boundary that restricts lateral transport and exerts comparatively limited energetic control over crystallographic orientation. These trends in adsorption behaviors provide an explanation for the experimentally observed angular distribution of nanowires relative to the template axis. Since amorphous SiN substrate is the dominant energetic stabilizer during nucleation, there is no strong driving force compelling the evolving crystallite to elongate parallel to the PMMA sidewalls. An additional notable outcome is that the absolute absorptive strength between Te and a-SiN or PMMA decreases with increasing Te size (Figure S10 b, c). When adsorption energies are normalized per Te length for periodic atomic chains, the difference between Te-SiN and Te-PMMA interactions decreases from the single-chain to the multi-chain nanowire model. This suggests that, as the Te crystal grows thicker, the energetic benefit of adhering to a particular boundary becomes less decisive. In other words, interfacial adhesion may be most consequential during nucleation and early growth, while later-stage growth is progressively governed by internal structural relaxation and strong Te-Te interactions (whose adsorption energy is approximately -0.4 eV/nm). This size-dependence of interfacial affinity is consistent with the experimentally observed correlation between template width and angular spread. As the confined crystal thickens, boundary adhesion becomes less dominant relative to internal Te-Te cohesion and structural relaxation. As a result, wider templates that accommodate thicker nanowires lead to reduced contribution of boundary adhesion and allow a wider range of nanowire orientations with respect to PMMA sidewalls. On the other hand, while adsorption energetics alone do not determine the full local strain tensor during crystallization, they do show that the substrate can interact strongly enough with small Te nuclei to perturb their structure and plausibly contribute to the constrained environments considered in the strain-torsion analysis.

## Computational Methodology

### I. 4DSTEM Simulation

To generate atomic model of a twisted Te nanowire, we start from an ideal Te crystal with lattice parameters of $a = b = 4.6$ Å and $c = 5.9$ Å[7]. Based on the ideal structure, an elongated supercell is constructed along c-axis. The resulting nanowire has dimension of $6 \times 15 \times 600$ nm$^3$. A helical twist is then applied to the atomic structure with a twist rate of 0.144° nm$^{-1}$. The twist is applied about the *c*-axis, with the center of mass the structure taken as the rotation center. Simulated 4DSTEM dataset of the generated Te twisted nanowire were generated using the abTEM package based on multislice formalism[4]. An electron probe with an accelerating voltage of 300 kV and a semi-convergence angle of 0.5 mrad was raster-scanned over the sample in the *x*-*y* plane with a step size of 24 nm. At each probe position, the far-field diffraction intensity was recorded using a pixelated detector with a maximum collection angle of 30 mrad. For multislice simulation, slice thickness of 1.6 Å is used, and thermal diffuse scattering was included via 8 frozen-phonon configurations. This simulation was performed under infinite-dose condition, for focusing on the helical structure. The final 4DSTEM dataset has dimensions of $1 \times 20 \times 250 \times 250$ after interpolation. Note that a one-dimensional scan was performed to reduce computational cost.

## II. DFT Calculations

Adsorption calculations were performed within the same DFT framework described in the main manuscript (VASP 6.4.1, PAW–PBE, 600 eV plane-wave cutoff, Gaussian smearing 0.03 eV, and SCF energy convergence of $1 \times 10^{-6}$ eV). Unless noted below, computational settings (electronic minimization parameters, smearing, and cutoff) were kept consistent across all adsorption calculations to enable direct comparison of adsorption energies.

Three Te building block configurations were constructed from the relaxed bulk trigonal Te structure: isolated Te dimer; single periodic Te helical chain oriented along Te [0001]; and minimal periodic nanowire model comprising three Te helical chains. Atomic positions of the isolated Te configurations were relaxed prior to adsorption calculations. Four substrates representative of the experimental environment were considered: isotactic PMMA, syndiotactic PMMA, H-terminated amorphous SiN, and OH-terminated amorphous SiN. PMMA was represented by a single polymer strand periodic along its chain direction, with vacuum in the transverse directions. Amorphous SiN was generated from a relaxed amorphous $Si_3N_4$ bulk model. Surface slabs were created by cleaving the bulk and passivating dangling bonds with either hydrogen or hydroxyl groups to form H-terminated and OH-terminated surfaces. To achieve consistent evaluation of adsorption energy, commensurate supercells were constructed such that one lattice dimension of each substrate supercell was an integer multiple of the corresponding Te periodic repeat length. For PMMA, Te periodicity was aligned with the polymer chain direction. For amorphous SiN slabs, which are periodic in-plane, Te periodicity was aligned along an in-plane direction of the slab. To minimize interactions between periodic images, a vacuum space of at least 15 Å was added in the non-periodic directions. Adsorbate-substrate geometries were relaxed for each Te-substrate complex to obtain energy minimized adsorption configurations. For all PMMA-containing supercells (adsorbed and isolated components), a Γ-centered 1×1×5 k-point mesh was used, with the 5-point sampling along the polymer periodic direction. For all amorphous SiN slab supercells (adsorbed and isolated components), a Γ-centered 3×5×1 k-point mesh was used to sample the two in-plane periodic directions of the slab. Adsorption energies were computed as:

$$E_{\mathrm{ads}} = E_{\mathrm{Te+sub}} - E_{\mathrm{Te}} - E_{\mathrm{sub}}$$

where $E_{\mathrm{Te+sub}}$ is the total energy of the relaxed Te-adsorbed substrate in the adsorption supercell, and $E_{\mathrm{Te}}$ and $E_{\mathrm{sub}}$ are the total energies of the relaxed, isolated Te configuration and clean substrates. For periodic Te chain models, adsorption energies were reported as energy per unit Te length by dividing $E_{\mathrm{ads}}$ by the Te periodic length within the supercell.